\documentclass[aip,preprint]{revtex4-1}
\usepackage{graphicx}
\usepackage{color}

\draft 

\begin{document}


\title{Experimental evidence of the non-diffusive avalanche-like electron heat transport events and their dynamical interaction with the shear flow structure} 

\author{Minjun J. Choi}
\email[]{mjchoi@nfri.re.kr}
\affiliation{National Fusion Research Institute, Daejeon 34133, Korea}
\author{Hogun Jhang}
\affiliation{National Fusion Research Institute, Daejeon 34133, Korea}
\author{J.-M. Kwon}
\affiliation{National Fusion Research Institute, Daejeon 34133, Korea}
\author{J. Chung}
\affiliation{National Fusion Research Institute, Daejeon 34133, Korea}
\author{M. Woo}
\affiliation{National Fusion Research Institute, Daejeon 34133, Korea}
\author{L. Qi}
\affiliation{National Fusion Research Institute, Daejeon 34133, Korea}
\author{S. H. Ko}
\affiliation{National Fusion Research Institute, Daejeon 34133, Korea}
\author{T. S. Hahm}
\affiliation{Seoul National University, Seoul 08826, Korea}
\author{H. K. Park}
\affiliation{Ulsan National Institute of Science and Technology, Ulsan 44919, Korea}
\author{H.-S. Kim}
\affiliation{National Fusion Research Institute, Daejeon 34133, Korea}
\author{J. S. Kang}
\affiliation{National Fusion Research Institute, Daejeon 34133, Korea}
\author{J. Lee}
\affiliation{National Fusion Research Institute, Daejeon 34133, Korea}
\author{M. Kim}
\affiliation{National Fusion Research Institute, Daejeon 34133, Korea}
\author{G. S. Yun}
\affiliation{Pohang University of Science and Technology, Pohang, Gyungbuk 37673, Korea}
\author{the KSTAR team}


\date{\today}

\begin{abstract}
We present experimental observations suggesting that the non-diffusive avalanche-like events are a prevalent and universal process of the electron turbulent heat transport in tokamak core plasmas. 
They are observed in the low confinement mode and the weak internal transport barrier tokamak plasmas in the absence of magnetohydrodynamic instabilities.
In addition, the electron temperature profile corrugation, which indicates the existence of the $E \times B$ shear flow layers, is clearly demonstrated as well as their dynamical interaction with the avalanche-like events.
The measured width of the profile corrugation is around $45\rho_i$, which implies the mesoscale nature of the structure.

\end{abstract}


\maketitle 


\section{Introduction}

Understanding dynamical processes leading to cross-field anomalous transport has been a central issue for more than four decades in magnetic fusion plasma research.
The prevailing paradigm at present is based on the local turbulent diffusive transport model~\cite{LiewerNF1985, HortonRMP1999, CarrerasIEEE1997} (far exceeding the transport level predicted by the neoclassical transport theory~\cite{HazeltineRMP1976}) self-regulated by turbulence generated mesoscale zonal flows~\cite{DiamondPPCF2005}.
A drawback of this local turbulent transport paradigm is that it does not explain the experimental observations in tokamaks exhibiting a strong deviation from the gyro-Bohm scaling~\cite{BudnyPoP2000, McKeeNF2002}.
To reconcile this discrepancy between the local turbulent transport theory and experimental observations, some non-diffusive transport processes, including turbulence spreading~\cite{GarbetNF1994, HahmPPCF2004, HahmJKPS2018} and/or avalanches~\cite{DiamondPoP1995, CarrerasIEEE1997, SanchezPPCF2015}, have been proposed.
These non-diffusive transport processes have been observed in many flux-driven fluid~\cite{CarrerasPoP1996, GarbetPoP1998, BenkaddaNF2001, MierPoP2006, TokunagaPoP2012} and gyrokinetic~\cite{IdomuraNF2009, McMillanPoP2009, KuNF2009, SarazinNF2010, DifPRE2010, JollietNF2012, DominskiPoP2015, WangNF2018, QiNF2019} simulations.
In particular, avalanches are near-ballistic radial transport events of heat accompanied by front propagation of fluctuation intensity, arising from nonlinear critical gradient dynamics.
They have been usually referred in relation with the paradigm of self-organized criticality (SOC)~\cite{BakPRL1987}. 
Once generated, they can reach to the edge region about one order of magnitude faster than the usual diffusion time, unless interrupted with a shear flow~\cite{DiamondPoP1995, CarrerasPoP1996, BenkaddaNF2001, IdomuraNF2009, DifPRE2010}.

Unfortunately, experimental observations of long range avalanches have been rare in tokamak experiments~\cite{PolitzerPRL2010}.
In addition, the avalanche dynamics in relation with the self-generated shear flow layers has not been demonstrated despite the shear flow layers were identified in turbulence coherence length measurements~\cite{DifPRL2015, HornungNF2017}.
This is partly due to the lack of an appropriate diagnostics which can measure temperature variations with a sufficient spatio-temporal resolution.
Another practical reason is a difficulty to suppress the magnetohydrodynamics (MHD) instabilities which can dominate the global transport and determine the profiles. 
In this paper, we show that non-diffusive avalanche-like transport events prevail in the electron heat transport channel in MHD-quiescent tokamak core plasmas, using advanced electron temperature ($T_\mathrm{e}$) imaging diagnostics~\cite{YunRSI2014}. 
We also provide observations of dynamical interaction between the avalanche-like events and self-generated shear flow layers.
The formation of the shear flow layers with a mesoscale width is demonstrated by a direct observation of the $T_\mathrm{e}$ profile corrugation.

\section{Experimental results}

\subsection{The avalanche-like events in the low confinement mode plasma}

\begin{figure}
\includegraphics[height=.55\textwidth, width=.7\textwidth]{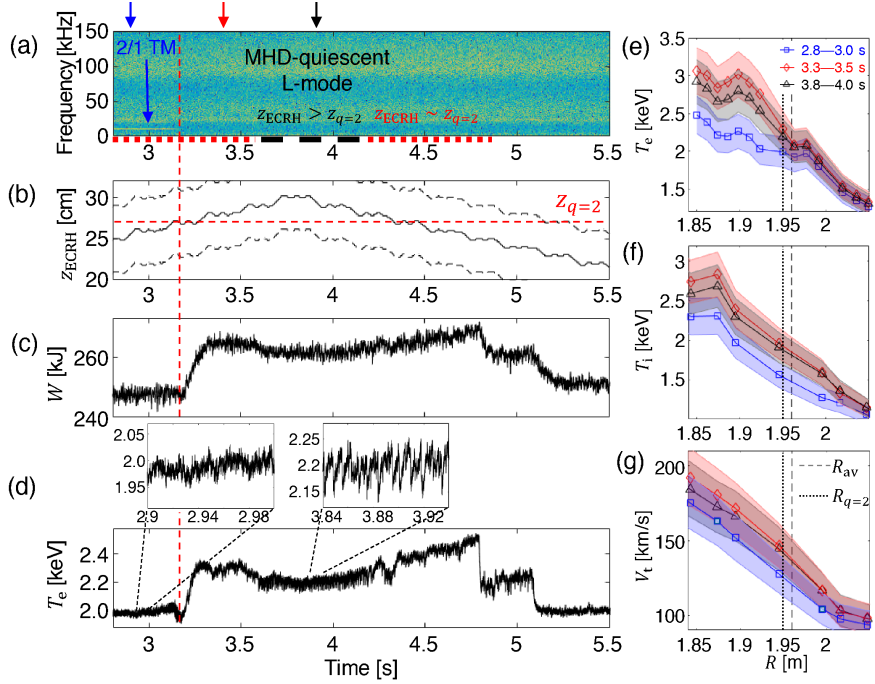}
\caption{\label{Loverall}
(Color online) (a) Power spectrogram of magnetic fluctuations. Bottom red dotted lines indicate the $z_\mathrm{ECRH} \sim z_{q=2}$ period and black dashed line indicates the $z_\mathrm{ECRH} > z_{q=2}$ period. (b) The vertical target position of ECRH ($z_\mathrm{ECRH}$, bold line) and effective deposition widths (black dashed lines). Red dashed horizontal line indicates the expected $q=2$ surface location ($z_{q=2}$). (c) The total stored energy in the plasma. (d) The electron temperature measurement near $R=1.95$~m just inside the avalanche initiation position ($R_\mathrm{av}$). (e)--(g) The average $T_\mathrm{e}$, $T_\mathrm{i}$, and $V_\mathrm{t}$ profiles for periods indicated by arrows of the corresponding colors in (a).
}
\end{figure}

The KSTAR~\cite{OhJKPS2018} discharge \#13728 is a limiter low confinement mode (L-mode) plasma with the major radius $R_0=1.8$~m, the minor radius $a\sim0.4$~m, the elongation $\kappa \sim 1.4$, the triangularity $\delta \sim 0.3$, the toroidal field $B_\mathrm{T}=3.0$~T, the plasma current $I_\mathrm{p} = 500$~kA, and a monotonic safety factor $q$ profile ($q_{95}\sim 7$ at the 95\% normalized poloidal flux).
The 4 MW of neutral beam power is applied to heat up the plasma. 
An $m/n=2/1$ tearing mode (TM) is shown to be destabilized in the plasma current ramp-up phase.
Power spectrogram of magnetic fluctuations measured by a Mirnov coil~\cite{BakRSI2004} is shown in figure~\ref{Loverall}(a).
The $\sim10$~kHz signal shown in figure~\ref{Loverall}(a) originates from the TM which is stabilized by applying the 1 MW electron cyclotron resonance heating (ECRH). 
As shown in figure~\ref{Loverall}(b), the vertical target position of ECRH ($z_\mathrm{ECRH}$) has been swept along the resonance layer located at $R_\mathrm{ECRH} \sim 1.8$~m. 
The TM signal disappears around $t=3.17$~s when $z_\mathrm{ECRH}$ is expected to be close to the $z$ position of the $q=2$ flux surface at $R=R_\mathrm{ECRH}$.
The confinement is improved up to $10\%$ after the TM suppression.
The profiles of $T_\mathrm{e}$, the ion temperature ($T_\mathrm{i}$), and the toroidal velocity ($V_\mathrm{t}$) are shown in figure~\ref{Loverall}(e)--(g). 
The $T_\mathrm{e}$ profile is measured by the electron cyclotron emission (ECE) diagnostics~\cite{KogiRSI2010}, while the $T_\mathrm{i}$ and $V_\mathrm{t}$ profiles are estimated from the charge exchange spectroscopy (CES) diagnostics using the carbon impurity~\cite{KoIEEE2010}. 
After the TM suppression, the fast electron heat transport events start to occur as shown in figure~\ref{Loverall}(d). 
These events are more clearly detected in the $z_\mathrm{ECRH}> z_{q=2}$ period, while in the $z_\mathrm{ECRH} \sim z_{q=2}$ period the high frequency coupled modes of $T_\mathrm{e}$ fluctuations appear, affecting dynamics of the transport events.
Characteristics of the high frequency modes are under more investigations, and in this paper we focus on the events observed in the $z_\mathrm{ECRH}> z_{q=2}$. 

\begin{figure}
\includegraphics[height=0.35\textwidth, width=0.48\textwidth]{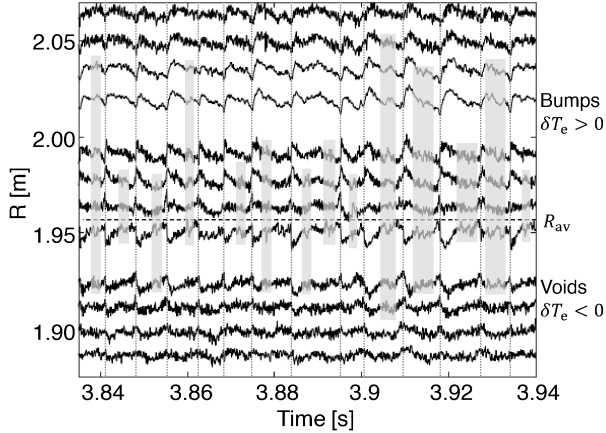}
\caption{\label{Ltrace}
(Color online) The rescaled $\tilde{T}_\mathrm{e}$ measurements at different radial locations (y-axis). Bumps ($\delta T_\mathrm{e} > 0$) propagate outwards in $R > R_\mathrm{av}$ (downhill) and voids ($\delta T_\mathrm{e} < 0$) propagate inwards in $R < R_\mathrm{av}$ (uphill).
}
\end{figure}

A detailed investigation of the $T_\mathrm{e}$ fluctuations over a broad radial region shows that the events are non-diffusive avalanche-like transport processes featuring different sizes and triggered near $R_\mathrm{av}\sim1.96$~m which is close to the q=2 flux surface $R_{q=2}\sim1.95$~m. 
Figure~\ref{Ltrace} shows the normalized $T_\mathrm{e}$ fluctuations ($\tilde{T}_\mathrm{e} \equiv (T_\mathrm{e} - \langle T_\mathrm{e} \rangle)/\langle T_\mathrm{e} \rangle$ where $ \langle ~ \rangle$ means a time average) measured at various radial positions (y-axis). 
$\tilde{T}_\mathrm{e}$ is filtered by a 5~kHz low-pass filter and rescaled to be in the $-1$ to $1$ range. 
Voids ($\delta T_\mathrm{e} = T_\mathrm{e} - \langle T_\mathrm{e} \rangle < 0$) and bumps ($\delta T_\mathrm{e} > 0$), propagating inwards and outwards from $R_\mathrm{av}\sim1.96$~m, respectively, are clearly observed.
Vertical dashed lines indicate the long range events whose large heat pulses propagate to the plasma boundary.
In addition to these large events, and there are also smaller ones highlighted by grey boxes. 
The smaller heat pulses propagate less than those of the large ones, implying a different radial transport scale.
The propagation speed of a heat pulse can be measured for the large event.
It is about $90~\mathrm{m/s}\sim 0.12 \left( \frac{\rho_\mathrm{s}}{a} \right) C_\mathrm{s}$, a fraction of the diamagnetic velocity~\cite{SarazinNF2011} where $\rho_\mathrm{s}$ is the sound Larmor radius and $C_\mathrm{s}$ is the sound speed. 
Therefore, the escaping time is estimated to be $t_\mathrm{sec} \sim 5$~msec, which is about 10 times faster than the energy confinement time. 
In the absence of MHD instabilities, these non-diffusive fast transport events are to be driven by turbulence. 
The heat avalanche has been proposed as a process for a system to relax perturbations in the context of the self-organized criticality (SOC) paradigm near marginal stability~\cite{BakPRL1987, DiamondPoP1995}.
The pair creation of a void and a bump and their respective upward and downward propagation can be interpreted as the joint reflection symmetry (JRS) which is a property expected in a SOC system~\cite{HwaPRA1992,DiamondPoP1995}.

\begin{figure}
\includegraphics[height=0.3\textwidth, width=0.35\textwidth]{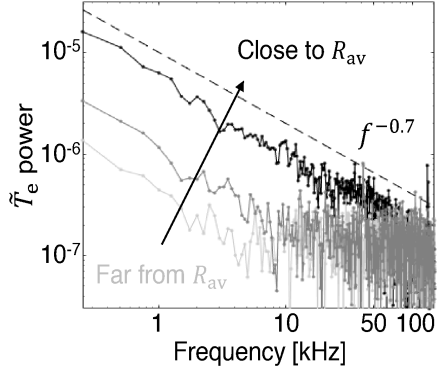}
\caption{\label{Lspectra} The power spectra of the $T_\mathrm{e}$ fluctuations. Black, grey, and light grey spectra are obtained by pairs of two adjacent ECEI channels on the same flux surface. 
}
\end{figure}

A further investigation on characteristics of $T_\mathrm{e}$ fluctuations corroborates that the observed events are non-diffusive and avalanche-like and they are the prevalent electron heat transport process in this plasma.
Figure~\ref{Lspectra} shows cross power spectra measured by pairs of two adjacent channels of electron cyclotron emission imaging (ECEI) diagnostics~\cite{YunRSI2014} during $t=$~3.7--4.0~s. 
The cross power spectrum near $R_\mathrm{av}$ where the avalanche-like event initiates has the largest fluctuation power with the power-law behavior, $S(f) \propto f^{-0.7}$, over 0--75~kHz frequency band (black).
As going far from $R_\mathrm{av}$ towards the center, the fluctuation powers of spectra are reduced while preserving the same spectral power-law behavior with an almost identical exponent (grey and light grey).
The power-law spectrum of event sizes implies that the transport events can occur in various scales, and rarity of the large size events represents the intrinsically intermittent nature of the large avalanche~\cite{HahmJKPS2018}.
The power-law spectrum is one of the most fundamental characteristics of an SOC system whose transport is likely to be governed by the avalanching process~\cite{HahmJKPS2018}.
A small deviation of the power-law exponent from $-1$ may result from the finite subdominant diffusion~\cite{MierPoP2008, SanchezPPCF2015}.
The observed $f^{-0.7}$ power-law spectra of $\delta T_\mathrm{e}$ reflect the non-diffusive and avalanche-like characteristics of electron heat transport in this plasma.
In accord with the power-law spectra, the Hurst exponent ($H$) measurement of the $T_\mathrm{e}$ fluctuation near $R_\mathrm{av}$ shows a long range temporal correlation ($H \sim 0.75 > 0.5$) representing a self-similar and persistent characteristics as observed in other avalanching systems~\cite{CarrerasPRL1998, PolitzerPoP2002, MierPoP2006, TokunagaPoP2012}. 
The Hurst exponent is calculated using the rescaled range statistics (R/S)~\cite{CarrerasPRL1998} for the time lag range of $0.1 < \tau < 1$~ms which corresponds to the frequency range of the power-law spectrum.

\begin{figure}
\includegraphics[height=0.3\textwidth, width=0.58\textwidth]{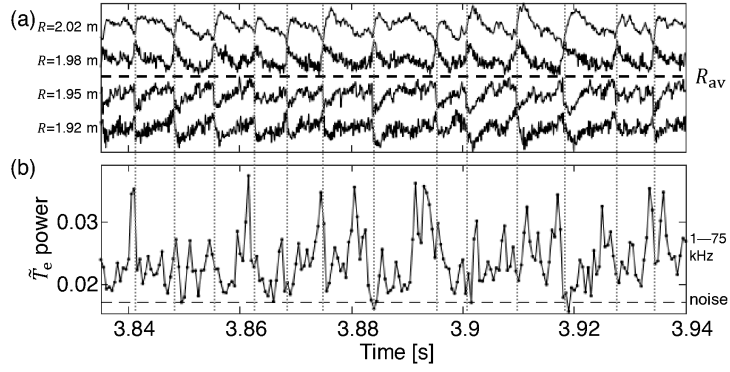}
\caption{\label{Lactivities} (a) The rescaled $\tilde{T}_\mathrm{e}$ at different $R$. (b) Fast temporal evolution of the $\tilde{T}_\mathrm{e}$ power near $R_\mathrm{av}$ in the 1--75~kHz frequency band. 
}
\end{figure}

Although an identification of the trigger mechanism for the avalanche-like event is beyond the scope of this paper, there is an indication that $\nabla T_\mathrm{e}$ is closely related to activities of the avalanche-like events. 
Activities of the avalanche-like events are quantified by the root mean square of the cross correlation, i.e. the cross power, between two $\tilde{T}_\mathrm{e}$ signals near $R_\mathrm{av}$ to reduce noise contributions.
The signals are first filtered by a 1--75~kHz band-pass filter to measure activities of small amplitude events. 
The 0--1~kHz range is excluded to avoid any possible errors from secular movements of the plasma (10--100~ms time scale).
As shown in figure~\ref{Lactivities}, activities clearly decrease when $|\nabla T_\mathrm{e}|$ at $R_\mathrm{av}$ ($T_\mathrm{e}$ difference between inside and outside of $R_\mathrm{av}$) collapses significantly due to large amplitude events. 
On the other hand, the long range and large amplitude events occur when activities of the small events and $|\nabla T_\mathrm{e}|$ are sufficiently large. 

\subsection{The avalanche-like events in the weak internal transport barrier plasma}

\begin{figure}
\centering
\includegraphics[height=0.5\textwidth, width=0.7\textwidth]{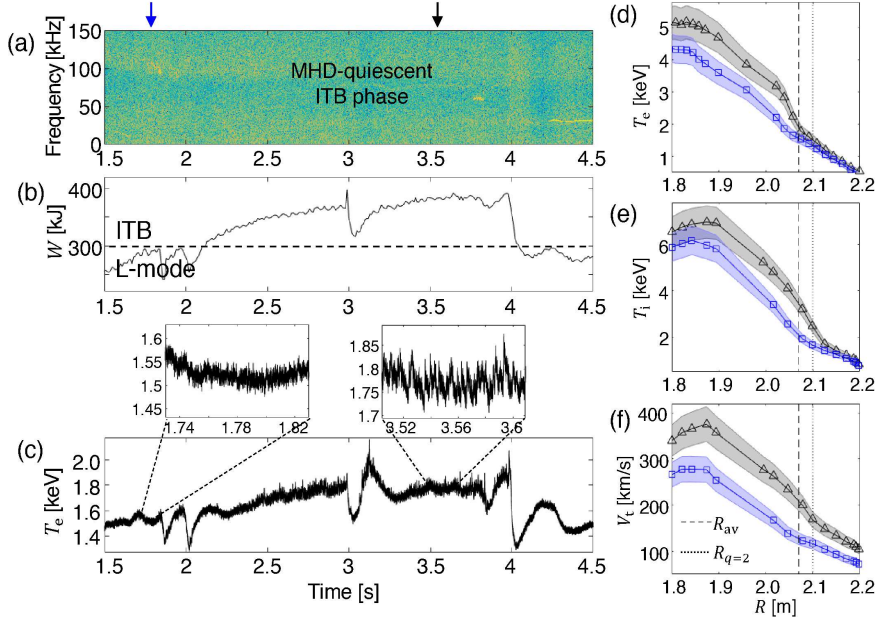}
\caption{(Color online) (a) Power spectrogram of magnetic fluctuations, (b) the stored energy, and (c) $T_\mathrm{e}$ measurement near at $R=2.1$~m. (d)--(f) The average $T_\mathrm{e}$, $T_\mathrm{i}$, and $V_\mathrm{t}$ profiles for periods indicated by arrows of the corresponding colors in (a). }
\label{Ioverall}
\end{figure}

The prevalence of avalanches can be a universal phenomenon as a plasma remains in marginally stable state~\cite{SanchezPPCF2015}.
This is a way for a magnetized plasma to relax quickly by expelling accumulated heat in the core from external heating.
Indeed, the similar avalanche-like electron heat transport event is observed in the plasma with a weak internal transport barrier (ITB), which possibly implies universality of the avalanche-like events in different tokamak confinement regimes.
Figure~\ref{Ioverall} shows time evolution of some characteristic plasma parameters of the KSTAR discharge \#17245~\cite{ChungNF2017}.
It is also a limiter plasma with a monotonic $q$ profile as the aforementioned L-mode discharge but with different $B_\mathrm{T}=2.7$~T, $I_\mathrm{p}=600$~kA, and $q_{95}\sim5.6$.
The 4.5 MW neutral beam (NB) power is injected at $t=0.5$~sec, i.e. in the early phase of the discharge to make an ITB.
The total stored energy shown in figure~\ref{Ioverall}(b) increases steadily with the formation of the ITB.
The maximum stored energy is about 390~kJ which is comparable to that of the typical KSTAR H-mode discharge.
Sudden decreases of the stored energy at 1.85, 2.0, 3.0, and 4.0 sec are due to NB blips which eventually cause the destruction of the ITB.  
Figures~\ref{Ioverall}(d)--(f) show the $T_\mathrm{e}$, $T_\mathrm{i}$, and $V_\mathrm{t}$ profiles demonstrating the formation of the ITBs.
Note that the ion and electron ITB foots are not exactly same, but the different central peak positions of $T_\mathrm{e}$ and $T_\mathrm{i}$ profiles might imply a few cm systematic error in channel positions in the profile diagnostics. 

During the ITB phase, there is a long period when MHD activities are quiescent.
It is this MHD-quiescent period that the avalanche-like events are clearly observed.
Figure~\ref{Ioverall}(a) shows power spectrogram of magnetic fluctuations and figure~\ref{Ioverall}(c) does the $T_\mathrm{e}$ measurement outside the avalanche-like event initiation position $R_\mathrm{av} \approx 2.07$~m.
Except for the unidentified weak 60~kHz signal, the spectrogram does not show any noticeable MHD signature.
The 30~kHz signal in the L-mode phase after the destruction of the ITB is identified as an $n=2$ tearing mode~\cite{ChungNF2017}. 

\begin{figure}
\centering
\includegraphics[height=0.3\textwidth,width=0.58\textwidth]{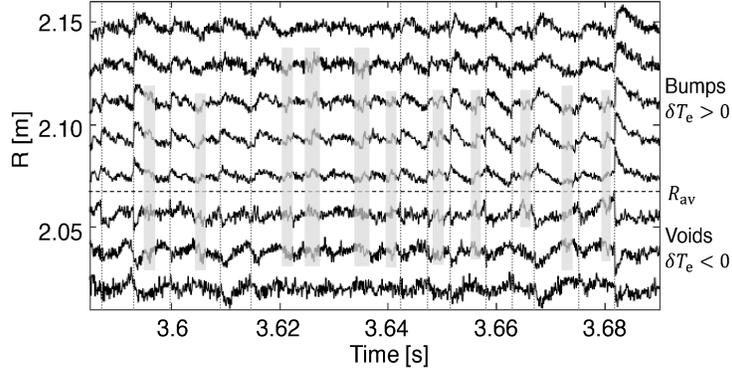}
\caption{The rescaled $\tilde{T}_\mathrm{e}$ measurements at different radial locations (y-axis). }
\label{Itrace}
\end{figure}

The avalanche-like events found in the MHD-quiescent period of the ITB plasma possess all the important characteristics discussed for the L-mode avalanche-like events.
They are the fast propagation speed about a fraction of the diamagnetic velocity ($140~\mathrm{m/s}\sim 0.18 \left( \frac{\rho_\mathrm{s}}{a} \right) C_\mathrm{s}$), the joint reflection symmetry, the long range temporal correlation $H \sim 0.8 > 0.5$, and the intermittent nature of the large avalanche or the power-law spectrum. 
$T_\mathrm{e}$ measurements over a broad radial region clearly reveal the ballistic nature of the events as shown in figure~\ref{Itrace}. 
Again, vertical dashed lines indicate the long range large amplitude events, while some smaller events are highlighted by grey boxes. 
The typical propagation speed of the heat pulse is as high as 140~m/s, and bumps and voids propagate in the opposite direction (JRS) from $R_\mathrm{av}$. 
Interestingly, $R_\mathrm{av}$ is also found to be close to the $R_{q=2}$ position as well as the electron ITB foot position. 
The Hurst exponent of the $T_\mathrm{e}$ fluctuation near $R_\mathrm{av}$ is about $H\sim0.8$, and the $T_\mathrm{e}$ fluctuation spectrum shows the similar power-law as in the avalanching L-mode plasma. 
However, smaller events are more suppressed in the ITB plasma, which might be related to the existence of a transport barrier close to the $R_\mathrm{av}$. 
Nonetheless, the heat accumulated inside the transport barrier escapes via the large avalanche-like events. 
These observations may imply that a common non-diffusive transport mechanism governs both the L-mode and the ITB turbulent electron heat transport.

\begin{figure}
\centering
\includegraphics[height=0.3\textwidth,width=0.35\textwidth]{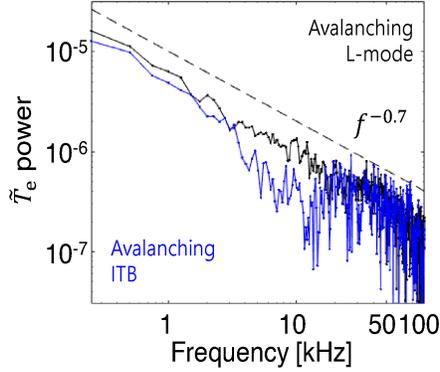}
\caption{(Color online) 
The power spectra of the $T_\mathrm{e}$ fluctuations for the ITB avalanching period (blue), compared with one obtained in the previous L-mode avalanching plasma (black). 
}\label{Ispectra}
\end{figure}

The avalanche-like events reported in this paper are clearly distinguished from MHD instabilities such as sawtooth, the tearing mode, the barrier localized mode (BLM), and the edge localized mode (ELM).
The avalanche-like events occur sporadically without any precursor or crash signature in magnetic or $T_\mathrm{e}$ fluctuation spectrum unlike the MHD instabilities.
In particular, BLM is often observed in the reversed shear (RS) $q$ ITB plasmas~\cite{TakejiPoP1998, ManickamNF1999}.
In KSTAR, BLM-like events are also observed when an ITB forms with the RS $q$ profile.

\subsection{Dynamical interaction between the avalanche-like events and the shear flow structure}

Recent studies suggest a hypothesis that the interplay between the avalanche and the self-generated shear flow layers may determine the turbulence transport scaling. 
It can be Bohm or worse-than-Bohm~\cite{CandyPRL2003, JollietNF2012} without the shear flow, while the gyro-Bohm scaling is recovered via a regulation of the avalanche activities by the self-organized mesoscale $E\times B$ shear flow layers~\cite{DifPRE2010, DifNF2017}.
There exist some experimental evidence that the $E \times B$ shear flow layers are present in tokamak plasmas~\cite{DifPRL2015, HornungNF2017}.
However, a ubiquitous signature for the presence of the shear layers, i.e. the temperature profile corrugation, and its dynamics have yet to be confirmed in experiments. 

\begin{figure}
\includegraphics[height=0.6\textwidth, width=0.68\textwidth]{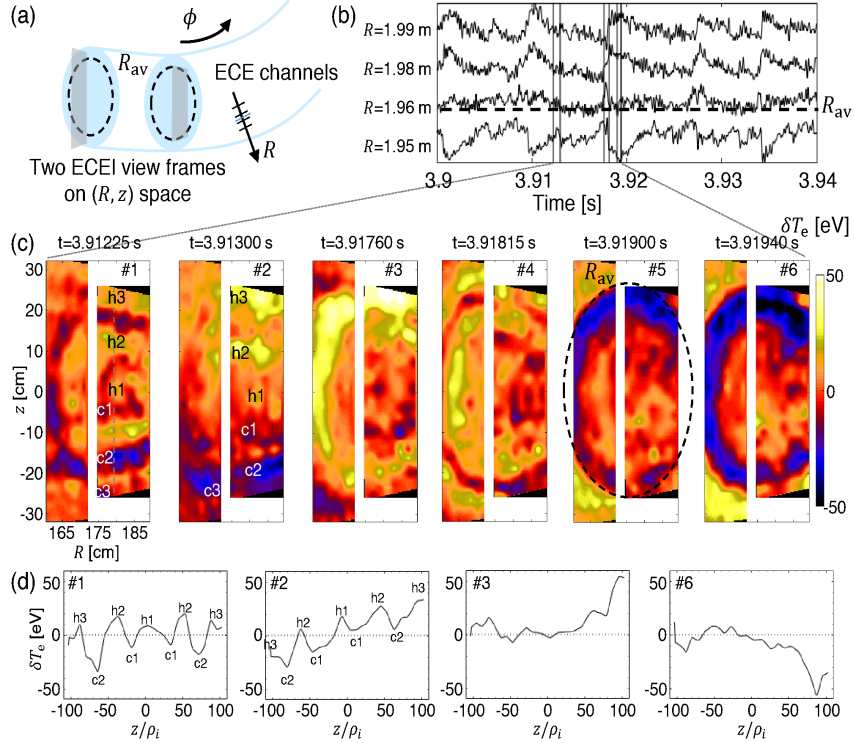}
\caption{\label{Lcorr}
(Color online) (a) The illustration of the diagnostics view. (b) The rescaled $\tilde{T}_\mathrm{e}$ at different $R$. (c) $\delta T_\mathrm{e}$ images during a single event. (d) Vertical cut of $\delta T_\mathrm{e}$ along the plasma center at the corresponding frame in (c). The h1--h3 and c1--c3 indicate positive and negative $\delta T_\mathrm{e}$ jets, respectively. 
}
\end{figure}

Interestingly, a careful investigation of the two-dimensional (2D) $T_\mathrm{e}$ variation in the avalanching L-mode plasma clearly captures the formation and destruction of the $T_\mathrm{e}$ profile corrugation.
Figure~\ref{Lcorr}(a) shows view frames of the ECEI diagnostics.
The ECEI data are filtered by a 3 kHz low-pass filter to reduce the noise while keeping $T_\mathrm{e}$ profile dynamics. 
After a preceding large event, the poloidally symmetric jet-like patterns appear as shown in the frame~\#1 in figure~\ref{Lcorr}(c) and its vertical cut in (d). 
The jet-like $\delta T_\mathrm{e}$ pattern implies that the $T_\mathrm{e}$ profile (and $\nabla T_\mathrm{e}$) is radially corrugated.
Spacing between the local maxima of $|\nabla T_\mathrm{e}|$ is roughly 10.8~cm, corresponding to $45\rho_i$ where $\rho_\mathrm{i}\sim0.24$~cm.
These symmetric $\delta T_\mathrm{e}$ jets are perturbed by a larger scale $m=1$ $\delta T_\mathrm{e}$ perturbation as shown in the frame~\#2.
The large $m=1$ perturbation has a peak at the top and a valley at the bottom (the $\sin \theta$ behavior). 
The $\delta T_\mathrm{e}$ jets move downwards about $10 \rho_\mathrm{i}$ within $750$~$\mu$s as its spacing expands about $5 \rho_\mathrm{i}$. 
As a side note, the $\delta T_\mathrm{e}$ polarity of the corrugation in the frame~\#1 and the $m=1$ perturbation in the frame \#2 are correlated each other and the opposite polarity of the corrugation with the $-\sin \theta$ perturbation are also observed. 
The $m=1$ perturbation becomes clear in between the frame \#2 and \#3 as the $\delta T_\mathrm{e}$ corrugation becomes blurred.
The local $T_\mathrm{e}$ near $R_\mathrm{av}$ (and the $\nabla T_\mathrm{e}$ across $R_\mathrm{av}$) increases significantly in the frame~\#3, and a long range large amplitude event occurs.
Inward (outward) propagation of a void (bump) in the $R<R_\mathrm{av}$ ($R>R_\mathrm{av}$) region is observed from the frame \#4 to \#6.

Observation of the $T_\mathrm{e}$ profile corrugation reflects a mesoscale transport regulation by the shear flow layers in turbulent avalanching plasmas.
It is noteworthy that the measured mesoscale spacing of the corrugations ($\sim 45 \rho_\mathrm{i}$) is similar with the values reported in the reference~\cite{DifNF2017}
($40 \pm 2 \rho_\mathrm{s} = 40 \pm 2 \sqrt{T_\mathrm{e}/T_\mathrm{i}} \rho_\mathrm{i} \approx 42 \pm 2 \rho_\mathrm{i}$),
though the KSTAR plasma parameters are different from the simulation conditions in the reference.
The $T_\mathrm{e}$ profile corrugation is destroyed with appearance of the $m=1$ perturbation, and then the long range large amplitude event could occur.
The $T_\mathrm{e}$ profile corrugation reforms in the wake of a large event~\cite{DifNF2017}.
An elucidation of the destruction mechanism of the $T_\mathrm{e}$ profile corrugation is beyond the scope of this paper.
One may speculate that it may involve a conversion of the zonal flow to the $m=1$ mode observed through a unknown mechanism.
Note that direct measurements of $E \times B$ shear flow layers were not available due to the lack of a fast flow diagnostics. 

\section{Discussion and summary}

Before closing this paper, we'd like to discuss several open issues to be addressed in future works.
Firstly, to understand the formation and destruction mechanisms of the $T_\mathrm{e}$ profile corrugation and the $E \times B$ shear flow layers, it is necessary to perform more cross validation researches. 
Again, we note the similarity of the spatial scale of the corrugation and shear flow layers from the KSTAR experiment and the previous works~\cite{DifNF2017}. 
This may hint a universal underlying physics controlling the mesoscale interaction of avalanches and shear flows, which is independent of detailed quantitative aspects of the plasma conditions or micro-physics.
Secondly, what determines the position of $R_\mathrm{av}$ where the event initiates is another important problem. 
In both the L-mode and ITB experiments, $R_\mathrm{av}$ is close to the $q=2$ flux surface.
Possible role of the rational surface in the avalanche dynamics may be studied further in $q$ scan experiments.
Note that the previous gyrokinetic simulations have reported the roles of rational flux surfaces in the destabilization of the trapped electron mode (TEM)~\cite{DominskiPoP2015} and enhanced electron transports at the low order rational flux surfaces~\cite{IdomuraJCP2016}.
Thirdly, since our study is only limited to the electron thermal transport, it is necessary to investigate other transport channels to compare their dynamics with the electron channel. 
Multi-field fluctuation measurements employing the various diagnostics (e.g. ECEI~\cite{YunRSI2014}, microwave imaging reflectometry~\cite{LeePPCF2018}, etc.) will be essential.
Also, numerical studies such as global gyrokinetic and gyrofluid simulations will provide a comprehensive physical picture of the overall transports.
Meanwhile, it is not very clear how the tokamak plasma which may be regarded as a strongly driven system, can exhibit the SOC-like dynamics, requiring a very slow perturbation (compared to the relaxation process). 
A further study is necessary to unveil this conundrum.

In summary, the avalanche-like electron heat transport events are observed in the L-mode and weak ITB tokamak core plasmas when the MHD instabilities are absent.
Experimental evidences supporting the non-diffusive and avalanche-like characteristics of the transport events are provided.
In addition, the existence of the $T_\mathrm{e}$ profile corrugation and their dynamical interaction with the avalanche-like events are clearly demonstrated. 
The measured width of the profile corrugation implies the mesoscale flow structure, limiting the size of avalanche-like events.
The long range avalanche-like events occur when the profile corrugation is destroyed. 




\begin{acknowledgments}
M.J. Choi and H. Jhang acknowledges helpful discussions with Dr. W. Wang, Prof. Y. Kishimoto, and Dr. K. Ida during the 4th UNIST-Kyoto workshop, 2018. 
J.-M. Kwon acknowledges helpful discussion with Dr. G. Dif-Pradalier.
This research was supported by Korea Ministry of Science and ICT under NFRI R\&D programs (NFRI-EN1801-9 and NFRI-EN1841-4).
\end{acknowledgments}


\begin{thebibliography}{47}%
\makeatletter
\providecommand \@ifxundefined [1]{%
 \@ifx{#1\undefined}
}%
\providecommand \@ifnum [1]{%
 \ifnum #1\expandafter \@firstoftwo
 \else \expandafter \@secondoftwo
 \fi
}%
\providecommand \@ifx [1]{%
 \ifx #1\expandafter \@firstoftwo
 \else \expandafter \@secondoftwo
 \fi
}%
\providecommand \natexlab [1]{#1}%
\providecommand \enquote  [1]{``#1''}%
\providecommand \bibnamefont  [1]{#1}%
\providecommand \bibfnamefont [1]{#1}%
\providecommand \citenamefont [1]{#1}%
\providecommand \href@noop [0]{\@secondoftwo}%
\providecommand \href [0]{\begingroup \@sanitize@url \@href}%
\providecommand \@href[1]{\@@startlink{#1}\@@href}%
\providecommand \@@href[1]{\endgroup#1\@@endlink}%
\providecommand \@sanitize@url [0]{\catcode `\\12\catcode `\$12\catcode
  `\&12\catcode `\#12\catcode `\^12\catcode `\_12\catcode `\%12\relax}%
\providecommand \@@startlink[1]{}%
\providecommand \@@endlink[0]{}%
\providecommand \url  [0]{\begingroup\@sanitize@url \@url }%
\providecommand \@url [1]{\endgroup\@href {#1}{\urlprefix }}%
\providecommand \urlprefix  [0]{URL }%
\providecommand \Eprint [0]{\href }%
\providecommand \doibase [0]{http://dx.doi.org/}%
\providecommand \selectlanguage [0]{\@gobble}%
\providecommand \bibinfo  [0]{\@secondoftwo}%
\providecommand \bibfield  [0]{\@secondoftwo}%
\providecommand \translation [1]{[#1]}%
\providecommand \BibitemOpen [0]{}%
\providecommand \bibitemStop [0]{}%
\providecommand \bibitemNoStop [0]{.\EOS\space}%
\providecommand \EOS [0]{\spacefactor3000\relax}%
\providecommand \BibitemShut  [1]{\csname bibitem#1\endcsname}%
\let\auto@bib@innerbib\@empty
\bibitem [{\citenamefont {Liewer}(1985)}]{LiewerNF1985}%
  \BibitemOpen
  \bibfield  {author} {\bibinfo {author} {\bibfnamefont {P.~C.}\ \bibnamefont
  {Liewer}},\ }\bibfield  {title} {\enquote {\bibinfo {title} {{Measurements of
  microturbulence in tokamaks and comparisons with theories of turbulence and
  anomalous transport}},}\ }\href@noop {} {\bibfield  {journal} {\bibinfo
  {journal} {Nuclear Fusion}\ }\textbf {\bibinfo {volume} {25}},\ \bibinfo
  {pages} {543--621} (\bibinfo {year} {1985})}\BibitemShut {NoStop}%
\bibitem [{\citenamefont {Horton}(1999)}]{HortonRMP1999}%
  \BibitemOpen
  \bibfield  {author} {\bibinfo {author} {\bibfnamefont {W.}~\bibnamefont
  {Horton}},\ }\bibfield  {title} {\enquote {\bibinfo {title} {{Drift waves and
  transport}},}\ }\href@noop {} {\bibfield  {journal} {\bibinfo  {journal}
  {Reviews of Modern Physics}\ }\textbf {\bibinfo {volume} {71}},\ \bibinfo
  {pages} {735--778} (\bibinfo {year} {1999})}\BibitemShut {NoStop}%
\bibitem [{\citenamefont {Carreras}(1997)}]{CarrerasIEEE1997}%
  \BibitemOpen
  \bibfield  {author} {\bibinfo {author} {\bibfnamefont {B.~A.}\ \bibnamefont
  {Carreras}},\ }\bibfield  {title} {\enquote {\bibinfo {title} {{Progress in
  anomalous transport research in toroidal magnetic confinement devices}},}\
  }\href@noop {} {\bibfield  {journal} {\bibinfo  {journal} {IEEE Transactions
  on Plasma Science}\ }\textbf {\bibinfo {volume} {25}},\ \bibinfo {pages}
  {1281--1321} (\bibinfo {year} {1997})}\BibitemShut {NoStop}%
\bibitem [{\citenamefont {Hinton}\ and\ \citenamefont
  {Hazeltine}(1976)}]{HazeltineRMP1976}%
  \BibitemOpen
  \bibfield  {author} {\bibinfo {author} {\bibfnamefont {F.~L.}\ \bibnamefont
  {Hinton}}\ and\ \bibinfo {author} {\bibfnamefont {R.~D.}\ \bibnamefont
  {Hazeltine}},\ }\bibfield  {title} {\enquote {\bibinfo {title} {{Theory of
  plasma transport in toroidal confinement systems}},}\ }\href@noop {}
  {\bibfield  {journal} {\bibinfo  {journal} {Reviews of Modern Physics}\
  }\textbf {\bibinfo {volume} {48}},\ \bibinfo {pages} {239--308} (\bibinfo
  {year} {1976})}\BibitemShut {NoStop}%
\bibitem [{\citenamefont {Diamond}\ \emph {et~al.}(2005)\citenamefont
  {Diamond}, \citenamefont {Itoh}, \citenamefont {Itoh},\ and\ \citenamefont
  {Hahm}}]{DiamondPPCF2005}%
  \BibitemOpen
  \bibfield  {author} {\bibinfo {author} {\bibfnamefont {P.~H.}\ \bibnamefont
  {Diamond}}, \bibinfo {author} {\bibfnamefont {S.~I.}\ \bibnamefont {Itoh}},
  \bibinfo {author} {\bibfnamefont {K.}~\bibnamefont {Itoh}}, \ and\ \bibinfo
  {author} {\bibfnamefont {T.~S.}\ \bibnamefont {Hahm}},\ }\bibfield  {title}
  {\enquote {\bibinfo {title} {{Zonal flows in plasma{\textemdash}a review}},}\
  }\href@noop {} {\bibfield  {journal} {\bibinfo  {journal} {Plasma Physics and
  Controlled Fusion}\ }\textbf {\bibinfo {volume} {47}},\ \bibinfo {pages}
  {R35--R161} (\bibinfo {year} {2005})}\BibitemShut {NoStop}%
\bibitem [{\citenamefont {Budny}\ \emph {et~al.}(2000)\citenamefont {Budny},
  \citenamefont {Ernst}, \citenamefont {Hahm}, \citenamefont {McCune},
  \citenamefont {Christiansen}, \citenamefont {Cordey}, \citenamefont {Gowers},
  \citenamefont {Guenther}, \citenamefont {Hawkes}, \citenamefont {Jarvis},
  \citenamefont {Stubberfield}, \citenamefont {Zastrow}, \citenamefont
  {Horton}, \citenamefont {Saibene}, \citenamefont {Sartori}, \citenamefont
  {Thomsen},\ and\ \citenamefont {von Hellermann}}]{BudnyPoP2000}%
  \BibitemOpen
  \bibfield  {author} {\bibinfo {author} {\bibfnamefont {R.~V.}\ \bibnamefont
  {Budny}}, \bibinfo {author} {\bibfnamefont {D.~R.}\ \bibnamefont {Ernst}},
  \bibinfo {author} {\bibfnamefont {T.~S.}\ \bibnamefont {Hahm}}, \bibinfo
  {author} {\bibfnamefont {D.~C.}\ \bibnamefont {McCune}}, \bibinfo {author}
  {\bibfnamefont {J.~P.}\ \bibnamefont {Christiansen}}, \bibinfo {author}
  {\bibfnamefont {J.~G.}\ \bibnamefont {Cordey}}, \bibinfo {author}
  {\bibfnamefont {C.~G.}\ \bibnamefont {Gowers}}, \bibinfo {author}
  {\bibfnamefont {K.}~\bibnamefont {Guenther}}, \bibinfo {author}
  {\bibfnamefont {N.}~\bibnamefont {Hawkes}}, \bibinfo {author} {\bibfnamefont
  {O.~N.}\ \bibnamefont {Jarvis}}, \bibinfo {author} {\bibfnamefont {P.~M.}\
  \bibnamefont {Stubberfield}}, \bibinfo {author} {\bibfnamefont {K.~D.}\
  \bibnamefont {Zastrow}}, \bibinfo {author} {\bibfnamefont {L.~D.}\
  \bibnamefont {Horton}}, \bibinfo {author} {\bibfnamefont {G.}~\bibnamefont
  {Saibene}}, \bibinfo {author} {\bibfnamefont {R.}~\bibnamefont {Sartori}},
  \bibinfo {author} {\bibfnamefont {K.}~\bibnamefont {Thomsen}}, \ and\
  \bibinfo {author} {\bibfnamefont {M.~G.}\ \bibnamefont {von Hellermann}},\
  }\bibfield  {title} {\enquote {\bibinfo {title} {{Local transport in Joint
  European Tokamak edge-localized, high-confinement mode plasmas with H, D, DT,
  and T isotopes}},}\ }\href@noop {} {\bibfield  {journal} {\bibinfo  {journal}
  {Physics of Plasmas}\ }\textbf {\bibinfo {volume} {7}},\ \bibinfo {pages}
  {5038--5050} (\bibinfo {year} {2000})}\BibitemShut {NoStop}%
\bibitem [{\citenamefont {McKee}\ \emph {et~al.}(2002)\citenamefont {McKee},
  \citenamefont {Petty}, \citenamefont {Waltz}, \citenamefont {Fenzi},
  \citenamefont {Fonck}, \citenamefont {Kinsey}, \citenamefont {Luce},
  \citenamefont {Burrell}, \citenamefont {Baker}, \citenamefont {Doyle},
  \citenamefont {Garbet}, \citenamefont {Moyer}, \citenamefont {Rettig},
  \citenamefont {Rhodes}, \citenamefont {Ross}, \citenamefont {Staebler},
  \citenamefont {Sydora},\ and\ \citenamefont {Wade}}]{McKeeNF2002}%
  \BibitemOpen
  \bibfield  {author} {\bibinfo {author} {\bibfnamefont {G.~R.}\ \bibnamefont
  {McKee}}, \bibinfo {author} {\bibfnamefont {C.~C.}\ \bibnamefont {Petty}},
  \bibinfo {author} {\bibfnamefont {R.~E.}\ \bibnamefont {Waltz}}, \bibinfo
  {author} {\bibfnamefont {C.}~\bibnamefont {Fenzi}}, \bibinfo {author}
  {\bibfnamefont {R.~J.}\ \bibnamefont {Fonck}}, \bibinfo {author}
  {\bibfnamefont {J.~E.}\ \bibnamefont {Kinsey}}, \bibinfo {author}
  {\bibfnamefont {T.~C.}\ \bibnamefont {Luce}}, \bibinfo {author}
  {\bibfnamefont {K.~H.}\ \bibnamefont {Burrell}}, \bibinfo {author}
  {\bibfnamefont {D.~R.}\ \bibnamefont {Baker}}, \bibinfo {author}
  {\bibfnamefont {E.~J.}\ \bibnamefont {Doyle}}, \bibinfo {author}
  {\bibfnamefont {X.}~\bibnamefont {Garbet}}, \bibinfo {author} {\bibfnamefont
  {R.~A.}\ \bibnamefont {Moyer}}, \bibinfo {author} {\bibfnamefont {C.~L.}\
  \bibnamefont {Rettig}}, \bibinfo {author} {\bibfnamefont {T.~L.}\
  \bibnamefont {Rhodes}}, \bibinfo {author} {\bibfnamefont {D.~W.}\
  \bibnamefont {Ross}}, \bibinfo {author} {\bibfnamefont {G.~M.}\ \bibnamefont
  {Staebler}}, \bibinfo {author} {\bibfnamefont {R.}~\bibnamefont {Sydora}}, \
  and\ \bibinfo {author} {\bibfnamefont {M.~R.}\ \bibnamefont {Wade}},\
  }\bibfield  {title} {\enquote {\bibinfo {title} {{Non-dimensional scaling of
  turbulence characteristics and turbulent diffusivity}},}\ }\href@noop {}
  {\bibfield  {journal} {\bibinfo  {journal} {Nuclear Fusion}\ }\textbf
  {\bibinfo {volume} {41}},\ \bibinfo {pages} {1235--1242} (\bibinfo {year}
  {2002})}\BibitemShut {NoStop}%
\bibitem [{\citenamefont {Garbet}\ \emph {et~al.}(1994)\citenamefont {Garbet},
  \citenamefont {Laurent}, \citenamefont {Samain},\ and\ \citenamefont
  {Chinardet}}]{GarbetNF1994}%
  \BibitemOpen
  \bibfield  {author} {\bibinfo {author} {\bibfnamefont {X.}~\bibnamefont
  {Garbet}}, \bibinfo {author} {\bibfnamefont {L.}~\bibnamefont {Laurent}},
  \bibinfo {author} {\bibfnamefont {A.}~\bibnamefont {Samain}}, \ and\ \bibinfo
  {author} {\bibfnamefont {J.}~\bibnamefont {Chinardet}},\ }\bibfield  {title}
  {\enquote {\bibinfo {title} {{Radial propagation of turbulence in
  tokamaks}},}\ }\href@noop {} {\bibfield  {journal} {\bibinfo  {journal}
  {Nuclear Fusion}\ }\textbf {\bibinfo {volume} {34}},\ \bibinfo {pages}
  {963--974} (\bibinfo {year} {1994})}\BibitemShut {NoStop}%
\bibitem [{\citenamefont {Hahm}\ \emph {et~al.}(2004)\citenamefont {Hahm},
  \citenamefont {Diamond}, \citenamefont {Lin}, \citenamefont {Itoh},\ and\
  \citenamefont {Itoh}}]{HahmPPCF2004}%
  \BibitemOpen
  \bibfield  {author} {\bibinfo {author} {\bibfnamefont {T.~S.}\ \bibnamefont
  {Hahm}}, \bibinfo {author} {\bibfnamefont {P.~H.}\ \bibnamefont {Diamond}},
  \bibinfo {author} {\bibfnamefont {Z.}~\bibnamefont {Lin}}, \bibinfo {author}
  {\bibfnamefont {K.}~\bibnamefont {Itoh}}, \ and\ \bibinfo {author}
  {\bibfnamefont {S.~I.}\ \bibnamefont {Itoh}},\ }\bibfield  {title} {\enquote
  {\bibinfo {title} {{Turbulence spreading into the linearly stable zone and
  transport scaling}},}\ }\href@noop {} {\bibfield  {journal} {\bibinfo
  {journal} {Plasma Physics and Controlled Fusion}\ }\textbf {\bibinfo {volume}
  {46}},\ \bibinfo {pages} {A323--A333} (\bibinfo {year} {2004})}\BibitemShut
  {NoStop}%
\bibitem [{\citenamefont {Hahm}\ and\ \citenamefont
  {Diamond}(2018)}]{HahmJKPS2018}%
  \BibitemOpen
  \bibfield  {author} {\bibinfo {author} {\bibfnamefont {T.~S.}\ \bibnamefont
  {Hahm}}\ and\ \bibinfo {author} {\bibfnamefont {P.~H.}\ \bibnamefont
  {Diamond}},\ }\bibfield  {title} {\enquote {\bibinfo {title} {{Mesoscopic
  Transport Events and the Breakdown of Fick{\textquoteright}s Law for
  Turbulent Fluxes }},}\ }\href@noop {} {\bibfield  {journal} {\bibinfo
  {journal} {Journal of the Korean Physical Society}\ }\textbf {\bibinfo
  {volume} {73}},\ \bibinfo {pages} {747--792} (\bibinfo {year}
  {2018})}\BibitemShut {NoStop}%
\bibitem [{\citenamefont {Diamond}\ and\ \citenamefont
  {Hahm}(1995)}]{DiamondPoP1995}%
  \BibitemOpen
  \bibfield  {author} {\bibinfo {author} {\bibfnamefont {P.~H.}\ \bibnamefont
  {Diamond}}\ and\ \bibinfo {author} {\bibfnamefont {T.~S.}\ \bibnamefont
  {Hahm}},\ }\bibfield  {title} {\enquote {\bibinfo {title} {{On the dynamics
  of turbulent transport near marginal stability}},}\ }\href@noop {} {\bibfield
   {journal} {\bibinfo  {journal} {Physics of Plasmas}\ }\textbf {\bibinfo
  {volume} {2}},\ \bibinfo {pages} {3640--3649} (\bibinfo {year}
  {1995})}\BibitemShut {NoStop}%
\bibitem [{\citenamefont {Sanchez}\ and\ \citenamefont
  {Newman}(2015)}]{SanchezPPCF2015}%
  \BibitemOpen
  \bibfield  {author} {\bibinfo {author} {\bibfnamefont {R.}~\bibnamefont
  {Sanchez}}\ and\ \bibinfo {author} {\bibfnamefont {D.~E.}\ \bibnamefont
  {Newman}},\ }\bibfield  {title} {\enquote {\bibinfo {title} {{Self-organized
  criticality and the dynamics of near-marginal turbulent transport in
  magnetically confined fusion plasmas}},}\ }\href@noop {} {\bibfield
  {journal} {\bibinfo  {journal} {Plasma Physics and Controlled Fusion}\
  }\textbf {\bibinfo {volume} {57}},\ \bibinfo {pages} {123002} (\bibinfo
  {year} {2015})}\BibitemShut {NoStop}%
\bibitem [{\citenamefont {Carreras}\ \emph {et~al.}(1996)\citenamefont
  {Carreras}, \citenamefont {Newman}, \citenamefont {Lynch},\ and\
  \citenamefont {Diamond}}]{CarrerasPoP1996}%
  \BibitemOpen
  \bibfield  {author} {\bibinfo {author} {\bibfnamefont {B.~A.}\ \bibnamefont
  {Carreras}}, \bibinfo {author} {\bibfnamefont {D.}~\bibnamefont {Newman}},
  \bibinfo {author} {\bibfnamefont {V.~E.}\ \bibnamefont {Lynch}}, \ and\
  \bibinfo {author} {\bibfnamefont {P.~H.}\ \bibnamefont {Diamond}},\
  }\bibfield  {title} {\enquote {\bibinfo {title} {{A model realization of
  self-organized criticality for plasma confinement}},}\ }\href@noop {}
  {\bibfield  {journal} {\bibinfo  {journal} {Physics of Plasmas}\ }\textbf
  {\bibinfo {volume} {3}},\ \bibinfo {pages} {2903--2911} (\bibinfo {year}
  {1996})}\BibitemShut {NoStop}%
\bibitem [{\citenamefont {Garbet}\ and\ \citenamefont
  {Waltz}(1998)}]{GarbetPoP1998}%
  \BibitemOpen
  \bibfield  {author} {\bibinfo {author} {\bibfnamefont {X.}~\bibnamefont
  {Garbet}}\ and\ \bibinfo {author} {\bibfnamefont {R.~E.}\ \bibnamefont
  {Waltz}},\ }\bibfield  {title} {\enquote {\bibinfo {title} {{Heat flux driven
  ion turbulence}},}\ }\href@noop {} {\bibfield  {journal} {\bibinfo  {journal}
  {Physics of Plasmas}\ }\textbf {\bibinfo {volume} {5}},\ \bibinfo {pages}
  {2836--2845} (\bibinfo {year} {1998})}\BibitemShut {NoStop}%
\bibitem [{\citenamefont {Benkadda}\ \emph {et~al.}(2001)\citenamefont
  {Benkadda}, \citenamefont {Beyer}, \citenamefont {Bian}, \citenamefont
  {Figarella}, \citenamefont {Garcia}, \citenamefont {Garbet}, \citenamefont
  {Ghendrih}, \citenamefont {Sarazin},\ and\ \citenamefont
  {Diamond}}]{BenkaddaNF2001}%
  \BibitemOpen
  \bibfield  {author} {\bibinfo {author} {\bibfnamefont {S.}~\bibnamefont
  {Benkadda}}, \bibinfo {author} {\bibfnamefont {P.}~\bibnamefont {Beyer}},
  \bibinfo {author} {\bibfnamefont {N.}~\bibnamefont {Bian}}, \bibinfo {author}
  {\bibfnamefont {C.}~\bibnamefont {Figarella}}, \bibinfo {author}
  {\bibfnamefont {O.}~\bibnamefont {Garcia}}, \bibinfo {author} {\bibfnamefont
  {X.}~\bibnamefont {Garbet}}, \bibinfo {author} {\bibfnamefont
  {P.}~\bibnamefont {Ghendrih}}, \bibinfo {author} {\bibfnamefont
  {Y.}~\bibnamefont {Sarazin}}, \ and\ \bibinfo {author} {\bibfnamefont
  {P.~H.}\ \bibnamefont {Diamond}},\ }\bibfield  {title} {\enquote {\bibinfo
  {title} {{Bursty transport in tokamak turbulence: Role of zonal flows and
  internal transport barriers}},}\ }\href@noop {} {\bibfield  {journal}
  {\bibinfo  {journal} {Nuclear Fusion}\ }\textbf {\bibinfo {volume} {41}},\
  \bibinfo {pages} {995--1001} (\bibinfo {year} {2001})}\BibitemShut {NoStop}%
\bibitem [{\citenamefont {Mier}, \citenamefont {Garc{\'\i}a},\ and\
  \citenamefont {Sanchez}(2006)}]{MierPoP2006}%
  \BibitemOpen
  \bibfield  {author} {\bibinfo {author} {\bibfnamefont {J.~A.}\ \bibnamefont
  {Mier}}, \bibinfo {author} {\bibfnamefont {L.}~\bibnamefont {Garc{\'\i}a}}, \
  and\ \bibinfo {author} {\bibfnamefont {R.}~\bibnamefont {Sanchez}},\
  }\bibfield  {title} {\enquote {\bibinfo {title} {{Study of the interaction
  between diffusive and avalanche-like transport in near-critical
  dissipative-trapped-electron-mode turbulence}},}\ }\href@noop {} {\bibfield
  {journal} {\bibinfo  {journal} {Physics of Plasmas}\ }\textbf {\bibinfo
  {volume} {13}},\ \bibinfo {pages} {102308} (\bibinfo {year}
  {2006})}\BibitemShut {NoStop}%
\bibitem [{\citenamefont {Tokunaga}\ \emph {et~al.}(2012)\citenamefont
  {Tokunaga}, \citenamefont {Jhang}, \citenamefont {Kim},\ and\ \citenamefont
  {Diamond}}]{TokunagaPoP2012}%
  \BibitemOpen
  \bibfield  {author} {\bibinfo {author} {\bibfnamefont {S.}~\bibnamefont
  {Tokunaga}}, \bibinfo {author} {\bibfnamefont {H.}~\bibnamefont {Jhang}},
  \bibinfo {author} {\bibfnamefont {S.~S.}\ \bibnamefont {Kim}}, \ and\
  \bibinfo {author} {\bibfnamefont {P.~H.}\ \bibnamefont {Diamond}},\
  }\bibfield  {title} {\enquote {\bibinfo {title} {{A statistical analysis of
  avalanching heat transport in stationary enhanced core confinement
  regimes}},}\ }\href@noop {} {\bibfield  {journal} {\bibinfo  {journal}
  {Physics of Plasmas}\ }\textbf {\bibinfo {volume} {19}},\ \bibinfo {pages}
  {092303--14} (\bibinfo {year} {2012})}\BibitemShut {NoStop}%
\bibitem [{\citenamefont {Idomura}\ \emph {et~al.}(2009)\citenamefont
  {Idomura}, \citenamefont {Urano}, \citenamefont {Aiba},\ and\ \citenamefont
  {Tokuda}}]{IdomuraNF2009}%
  \BibitemOpen
  \bibfield  {author} {\bibinfo {author} {\bibfnamefont {Y.}~\bibnamefont
  {Idomura}}, \bibinfo {author} {\bibfnamefont {H.}~\bibnamefont {Urano}},
  \bibinfo {author} {\bibfnamefont {N.}~\bibnamefont {Aiba}}, \ and\ \bibinfo
  {author} {\bibfnamefont {S.}~\bibnamefont {Tokuda}},\ }\bibfield  {title}
  {\enquote {\bibinfo {title} {{Study of ion turbulent transport and profile
  formations using global gyrokinetic full-f Vlasov simulation}},}\ }\href@noop
  {} {\bibfield  {journal} {\bibinfo  {journal} {Nuclear Fusion}\ }\textbf
  {\bibinfo {volume} {49}},\ \bibinfo {pages} {065029} (\bibinfo {year}
  {2009})}\BibitemShut {NoStop}%
\bibitem [{\citenamefont {McMillan}\ \emph {et~al.}(2009)\citenamefont
  {McMillan}, \citenamefont {Jolliet}, \citenamefont {Tran}, \citenamefont
  {Villard}, \citenamefont {Bottino},\ and\ \citenamefont
  {Angelino}}]{McMillanPoP2009}%
  \BibitemOpen
  \bibfield  {author} {\bibinfo {author} {\bibfnamefont {B.~F.}\ \bibnamefont
  {McMillan}}, \bibinfo {author} {\bibfnamefont {S.}~\bibnamefont {Jolliet}},
  \bibinfo {author} {\bibfnamefont {T.~M.}\ \bibnamefont {Tran}}, \bibinfo
  {author} {\bibfnamefont {L.}~\bibnamefont {Villard}}, \bibinfo {author}
  {\bibfnamefont {A.}~\bibnamefont {Bottino}}, \ and\ \bibinfo {author}
  {\bibfnamefont {P.}~\bibnamefont {Angelino}},\ }\bibfield  {title} {\enquote
  {\bibinfo {title} {{Avalanchelike bursts in global gyrokinetic
  simulations}},}\ }\href@noop {} {\bibfield  {journal} {\bibinfo  {journal}
  {Physics of Plasmas}\ }\textbf {\bibinfo {volume} {16}},\ \bibinfo {pages}
  {022310} (\bibinfo {year} {2009})}\BibitemShut {NoStop}%
\bibitem [{\citenamefont {Ku}, \citenamefont {Chang},\ and\ \citenamefont
  {Diamond}(2009)}]{KuNF2009}%
  \BibitemOpen
  \bibfield  {author} {\bibinfo {author} {\bibfnamefont {S.}~\bibnamefont
  {Ku}}, \bibinfo {author} {\bibfnamefont {C.~S.}\ \bibnamefont {Chang}}, \
  and\ \bibinfo {author} {\bibfnamefont {P.~H.}\ \bibnamefont {Diamond}},\
  }\bibfield  {title} {\enquote {\bibinfo {title} {{Full-f gyrokinetic particle
  simulation of centrally heated global ITG turbulence from magnetic axis to
  edge pedestal top in a realistic tokamak geometry}},}\ }\href@noop {}
  {\bibfield  {journal} {\bibinfo  {journal} {Nuclear Fusion}\ }\textbf
  {\bibinfo {volume} {49}},\ \bibinfo {pages} {115021} (\bibinfo {year}
  {2009})}\BibitemShut {NoStop}%
\bibitem [{\citenamefont {Sarazin}\ \emph {et~al.}(2010)\citenamefont
  {Sarazin}, \citenamefont {Grandgirard}, \citenamefont {Abiteboul},
  \citenamefont {Allfrey}, \citenamefont {Garbet}, \citenamefont {Ghendrih},
  \citenamefont {Latu}, \citenamefont {Strugarek},\ and\ \citenamefont
  {Dif-Pradalier}}]{SarazinNF2010}%
  \BibitemOpen
  \bibfield  {author} {\bibinfo {author} {\bibfnamefont {Y.}~\bibnamefont
  {Sarazin}}, \bibinfo {author} {\bibfnamefont {V.}~\bibnamefont
  {Grandgirard}}, \bibinfo {author} {\bibfnamefont {J.}~\bibnamefont
  {Abiteboul}}, \bibinfo {author} {\bibfnamefont {S.}~\bibnamefont {Allfrey}},
  \bibinfo {author} {\bibfnamefont {X.}~\bibnamefont {Garbet}}, \bibinfo
  {author} {\bibfnamefont {P.}~\bibnamefont {Ghendrih}}, \bibinfo {author}
  {\bibfnamefont {G.}~\bibnamefont {Latu}}, \bibinfo {author} {\bibfnamefont
  {A.}~\bibnamefont {Strugarek}}, \ and\ \bibinfo {author} {\bibfnamefont
  {G.}~\bibnamefont {Dif-Pradalier}},\ }\bibfield  {title} {\enquote {\bibinfo
  {title} {{Large scale dynamics in flux driven gyrokinetic turbulence}},}\
  }\href@noop {} {\bibfield  {journal} {\bibinfo  {journal} {Nuclear Fusion}\
  }\textbf {\bibinfo {volume} {50}},\ \bibinfo {pages} {054004} (\bibinfo
  {year} {2010})}\BibitemShut {NoStop}%
\bibitem [{\citenamefont {Dif-Pradalier}\ \emph {et~al.}(2010)\citenamefont
  {Dif-Pradalier}, \citenamefont {Diamond}, \citenamefont {Grandgirard},
  \citenamefont {Sarazin}, \citenamefont {Abiteboul}, \citenamefont {Garbet},
  \citenamefont {Ghendrih}, \citenamefont {Strugarek}, \citenamefont {Ku},\
  and\ \citenamefont {Chang}}]{DifPRE2010}%
  \BibitemOpen
  \bibfield  {author} {\bibinfo {author} {\bibfnamefont {G.}~\bibnamefont
  {Dif-Pradalier}}, \bibinfo {author} {\bibfnamefont {P.~H.}\ \bibnamefont
  {Diamond}}, \bibinfo {author} {\bibfnamefont {V.}~\bibnamefont
  {Grandgirard}}, \bibinfo {author} {\bibfnamefont {Y.}~\bibnamefont
  {Sarazin}}, \bibinfo {author} {\bibfnamefont {J.}~\bibnamefont {Abiteboul}},
  \bibinfo {author} {\bibfnamefont {X.}~\bibnamefont {Garbet}}, \bibinfo
  {author} {\bibfnamefont {P.}~\bibnamefont {Ghendrih}}, \bibinfo {author}
  {\bibfnamefont {A.}~\bibnamefont {Strugarek}}, \bibinfo {author}
  {\bibfnamefont {S.}~\bibnamefont {Ku}}, \ and\ \bibinfo {author}
  {\bibfnamefont {C.~S.}\ \bibnamefont {Chang}},\ }\bibfield  {title} {\enquote
  {\bibinfo {title} {{On the validity of the local diffusive paradigm in
  turbulent plasma transport}},}\ }\href@noop {} {\bibfield  {journal}
  {\bibinfo  {journal} {Physical Review E}\ }\textbf {\bibinfo {volume} {82}},\
  \bibinfo {pages} {025401(R)} (\bibinfo {year} {2010})}\BibitemShut {NoStop}%
\bibitem [{\citenamefont {Jolliet}\ and\ \citenamefont
  {Idomura}(2012)}]{JollietNF2012}%
  \BibitemOpen
  \bibfield  {author} {\bibinfo {author} {\bibfnamefont {S.}~\bibnamefont
  {Jolliet}}\ and\ \bibinfo {author} {\bibfnamefont {Y.}~\bibnamefont
  {Idomura}},\ }\bibfield  {title} {\enquote {\bibinfo {title} {{Plasma size
  scaling of avalanche-like heat transport in tokamaks}},}\ }\href@noop {}
  {\bibfield  {journal} {\bibinfo  {journal} {Nuclear Fusion}\ }\textbf
  {\bibinfo {volume} {52}},\ \bibinfo {pages} {023026} (\bibinfo {year}
  {2012})}\BibitemShut {NoStop}%
\bibitem [{\citenamefont {Dominski}\ \emph {et~al.}(2015)\citenamefont
  {Dominski}, \citenamefont {Brunner}, \citenamefont {G{\"o}rler},
  \citenamefont {Jenko}, \citenamefont {Told},\ and\ \citenamefont
  {Villard}}]{DominskiPoP2015}%
  \BibitemOpen
  \bibfield  {author} {\bibinfo {author} {\bibfnamefont {J.}~\bibnamefont
  {Dominski}}, \bibinfo {author} {\bibfnamefont {S.}~\bibnamefont {Brunner}},
  \bibinfo {author} {\bibfnamefont {T.}~\bibnamefont {G{\"o}rler}}, \bibinfo
  {author} {\bibfnamefont {F.}~\bibnamefont {Jenko}}, \bibinfo {author}
  {\bibfnamefont {D.}~\bibnamefont {Told}}, \ and\ \bibinfo {author}
  {\bibfnamefont {L.}~\bibnamefont {Villard}},\ }\bibfield  {title} {\enquote
  {\bibinfo {title} {{How non-adiabatic passing electron layers of linear
  microinstabilities affect turbulent transport}},}\ }\href@noop {} {\bibfield
  {journal} {\bibinfo  {journal} {Physics of Plasmas}\ }\textbf {\bibinfo
  {volume} {22}},\ \bibinfo {pages} {062303} (\bibinfo {year}
  {2015})}\BibitemShut {NoStop}%
\bibitem [{\citenamefont {Wang}\ \emph {et~al.}(2018)\citenamefont {Wang},
  \citenamefont {Kishimoto}, \citenamefont {Imadera}, \citenamefont {Li},\ and\
  \citenamefont {Wang}}]{WangNF2018}%
  \BibitemOpen
  \bibfield  {author} {\bibinfo {author} {\bibfnamefont {W.}~\bibnamefont
  {Wang}}, \bibinfo {author} {\bibfnamefont {Y.}~\bibnamefont {Kishimoto}},
  \bibinfo {author} {\bibfnamefont {K.}~\bibnamefont {Imadera}}, \bibinfo
  {author} {\bibfnamefont {J.~Q.}\ \bibnamefont {Li}}, \ and\ \bibinfo {author}
  {\bibfnamefont {Z.~X.}\ \bibnamefont {Wang}},\ }\bibfield  {title} {\enquote
  {\bibinfo {title} {{A mechanism for the formation and sustainment of the
  self-organized global profile and E $\times$ B staircase in tokamak
  plasmas}},}\ }\href@noop {} {\bibfield  {journal} {\bibinfo  {journal}
  {Nuclear Fusion}\ }\textbf {\bibinfo {volume} {58}},\ \bibinfo {pages}
  {056005} (\bibinfo {year} {2018})}\BibitemShut {NoStop}%
\bibitem [{\citenamefont {Qi}\ \emph {et~al.}(2019)\citenamefont {Qi},
  \citenamefont {Kwon}, \citenamefont {Hahm}, \citenamefont {Yi},\ and\
  \citenamefont {Choi}}]{QiNF2019}%
  \BibitemOpen
  \bibfield  {author} {\bibinfo {author} {\bibfnamefont {L.}~\bibnamefont
  {Qi}}, \bibinfo {author} {\bibfnamefont {J.-M.}\ \bibnamefont {Kwon}},
  \bibinfo {author} {\bibfnamefont {T.~S.}\ \bibnamefont {Hahm}}, \bibinfo
  {author} {\bibfnamefont {S.}~\bibnamefont {Yi}}, \ and\ \bibinfo {author}
  {\bibfnamefont {M.~J.}\ \bibnamefont {Choi}},\ }\bibfield  {title} {\enquote
  {\bibinfo {title} {{Characteristics of trapped electron transport, zonal flow
  staircase, turbulence fluctuation spectra in elongated tokamak plasmas}},}\
  }\href@noop {} {\bibfield  {journal} {\bibinfo  {journal} {Nuclear Fusion}\
  }\textbf {\bibinfo {volume} {59}},\ \bibinfo {pages} {026013} (\bibinfo
  {year} {2019})}\BibitemShut {NoStop}%
\bibitem [{\citenamefont {Bak}, \citenamefont {Tang},\ and\ \citenamefont
  {Wiesenfeld}(1987)}]{BakPRL1987}%
  \BibitemOpen
  \bibfield  {author} {\bibinfo {author} {\bibfnamefont {P.}~\bibnamefont
  {Bak}}, \bibinfo {author} {\bibfnamefont {C.}~\bibnamefont {Tang}}, \ and\
  \bibinfo {author} {\bibfnamefont {K.}~\bibnamefont {Wiesenfeld}},\ }\bibfield
   {title} {\enquote {\bibinfo {title} {{Self-organized criticality: An
  explanation of the 1/ f noise}},}\ }\href@noop {} {\bibfield  {journal}
  {\bibinfo  {journal} {Physical Review Letters}\ }\textbf {\bibinfo {volume}
  {59}},\ \bibinfo {pages} {381--384} (\bibinfo {year} {1987})}\BibitemShut
  {NoStop}%
\bibitem [{\citenamefont {Politzer}(2000)}]{PolitzerPRL2010}%
  \BibitemOpen
  \bibfield  {author} {\bibinfo {author} {\bibfnamefont {P.~A.}\ \bibnamefont
  {Politzer}},\ }\bibfield  {title} {\enquote {\bibinfo {title} {{Observation
  of Avalanchelike Phenomena in a Magnetically Confined Plasma}},}\ }\href@noop
  {} {\bibfield  {journal} {\bibinfo  {journal} {Physical Review Letters}\
  }\textbf {\bibinfo {volume} {84}},\ \bibinfo {pages} {1192--1195} (\bibinfo
  {year} {2000})}\BibitemShut {NoStop}%
\bibitem [{\citenamefont {Dif-Pradalier}\ \emph {et~al.}(2015)\citenamefont
  {Dif-Pradalier}, \citenamefont {Hornung}, \citenamefont {Ghendrih},
  \citenamefont {Sarazin}, \citenamefont {Clairet}, \citenamefont {Vermare},
  \citenamefont {Diamond}, \citenamefont {Abiteboul}, \citenamefont
  {Cartier-Michaud}, \citenamefont {Ehrlacher}, \citenamefont {Esteve},
  \citenamefont {Garbet}, \citenamefont {Grandgirard}, \citenamefont {Gurcan},
  \citenamefont {Hennequin}, \citenamefont {Kosuga}, \citenamefont {Latu},
  \citenamefont {Maget}, \citenamefont {Morel}, \citenamefont {Norscini},
  \citenamefont {Sabot},\ and\ \citenamefont {Storelli}}]{DifPRL2015}%
  \BibitemOpen
  \bibfield  {author} {\bibinfo {author} {\bibfnamefont {G.}~\bibnamefont
  {Dif-Pradalier}}, \bibinfo {author} {\bibfnamefont {G.}~\bibnamefont
  {Hornung}}, \bibinfo {author} {\bibfnamefont {P.}~\bibnamefont {Ghendrih}},
  \bibinfo {author} {\bibfnamefont {Y.}~\bibnamefont {Sarazin}}, \bibinfo
  {author} {\bibfnamefont {F.}~\bibnamefont {Clairet}}, \bibinfo {author}
  {\bibfnamefont {L.}~\bibnamefont {Vermare}}, \bibinfo {author} {\bibfnamefont
  {P.~H.}\ \bibnamefont {Diamond}}, \bibinfo {author} {\bibfnamefont
  {J.}~\bibnamefont {Abiteboul}}, \bibinfo {author} {\bibfnamefont
  {T.}~\bibnamefont {Cartier-Michaud}}, \bibinfo {author} {\bibfnamefont
  {C.}~\bibnamefont {Ehrlacher}}, \bibinfo {author} {\bibfnamefont
  {D.}~\bibnamefont {Esteve}}, \bibinfo {author} {\bibfnamefont
  {X.}~\bibnamefont {Garbet}}, \bibinfo {author} {\bibfnamefont
  {V.}~\bibnamefont {Grandgirard}}, \bibinfo {author} {\bibfnamefont {O.~D.}\
  \bibnamefont {Gurcan}}, \bibinfo {author} {\bibfnamefont {P.}~\bibnamefont
  {Hennequin}}, \bibinfo {author} {\bibfnamefont {Y.}~\bibnamefont {Kosuga}},
  \bibinfo {author} {\bibfnamefont {G.}~\bibnamefont {Latu}}, \bibinfo {author}
  {\bibfnamefont {P.}~\bibnamefont {Maget}}, \bibinfo {author} {\bibfnamefont
  {P.}~\bibnamefont {Morel}}, \bibinfo {author} {\bibfnamefont
  {C.}~\bibnamefont {Norscini}}, \bibinfo {author} {\bibfnamefont
  {R.}~\bibnamefont {Sabot}}, \ and\ \bibinfo {author} {\bibfnamefont
  {A.}~\bibnamefont {Storelli}},\ }\bibfield  {title} {\enquote {\bibinfo
  {title} {{Finding the elusive ExB staircase in magnetized plasmas}},}\
  }\href@noop {} {\bibfield  {journal} {\bibinfo  {journal} {Physical Review
  Letters}\ }\textbf {\bibinfo {volume} {114}},\ \bibinfo {pages} {085004}
  (\bibinfo {year} {2015})}\BibitemShut {NoStop}%
\bibitem [{\citenamefont {Hornung}\ \emph {et~al.}(2017)\citenamefont
  {Hornung}, \citenamefont {Dif-Pradalier}, \citenamefont {Clairet},
  \citenamefont {Sarazin}, \citenamefont {Sabot}, \citenamefont {Hennequin},\
  and\ \citenamefont {Verdoolaege}}]{HornungNF2017}%
  \BibitemOpen
  \bibfield  {author} {\bibinfo {author} {\bibfnamefont {G.}~\bibnamefont
  {Hornung}}, \bibinfo {author} {\bibfnamefont {G.}~\bibnamefont
  {Dif-Pradalier}}, \bibinfo {author} {\bibfnamefont {F.}~\bibnamefont
  {Clairet}}, \bibinfo {author} {\bibfnamefont {Y.}~\bibnamefont {Sarazin}},
  \bibinfo {author} {\bibfnamefont {R.}~\bibnamefont {Sabot}}, \bibinfo
  {author} {\bibfnamefont {P.}~\bibnamefont {Hennequin}}, \ and\ \bibinfo
  {author} {\bibfnamefont {G.}~\bibnamefont {Verdoolaege}},\ }\bibfield
  {title} {\enquote {\bibinfo {title} {{$\mathbf{E}\times \mathbf{B}$
  staircases and barrier permeability in magnetised plasmas}},}\ }\href@noop {}
  {\bibfield  {journal} {\bibinfo  {journal} {Nuclear Fusion}\ }\textbf
  {\bibinfo {volume} {57}},\ \bibinfo {pages} {014006} (\bibinfo {year}
  {2017})}\BibitemShut {NoStop}%
\bibitem [{\citenamefont {Yun}\ \emph {et~al.}(2014)\citenamefont {Yun},
  \citenamefont {Lee}, \citenamefont {Choi}, \citenamefont {Lee}, \citenamefont
  {Kim}, \citenamefont {Leem}, \citenamefont {Nam}, \citenamefont {Choe},
  \citenamefont {Park}, \citenamefont {Park}, \citenamefont {Woo},
  \citenamefont {Kim}, \citenamefont {Domier}, \citenamefont {Luhmann},
  \citenamefont {Ito}, \citenamefont {Mase},\ and\ \citenamefont
  {Lee}}]{YunRSI2014}%
  \BibitemOpen
  \bibfield  {author} {\bibinfo {author} {\bibfnamefont {G.~S.}\ \bibnamefont
  {Yun}}, \bibinfo {author} {\bibfnamefont {W.}~\bibnamefont {Lee}}, \bibinfo
  {author} {\bibfnamefont {M.~J.}\ \bibnamefont {Choi}}, \bibinfo {author}
  {\bibfnamefont {J.}~\bibnamefont {Lee}}, \bibinfo {author} {\bibfnamefont
  {M.}~\bibnamefont {Kim}}, \bibinfo {author} {\bibfnamefont {J.}~\bibnamefont
  {Leem}}, \bibinfo {author} {\bibfnamefont {Y.}~\bibnamefont {Nam}}, \bibinfo
  {author} {\bibfnamefont {G.~H.}\ \bibnamefont {Choe}}, \bibinfo {author}
  {\bibfnamefont {H.~K.}\ \bibnamefont {Park}}, \bibinfo {author}
  {\bibfnamefont {H.}~\bibnamefont {Park}}, \bibinfo {author} {\bibfnamefont
  {D.~S.}\ \bibnamefont {Woo}}, \bibinfo {author} {\bibfnamefont {K.~W.}\
  \bibnamefont {Kim}}, \bibinfo {author} {\bibfnamefont {C.~W.}\ \bibnamefont
  {Domier}}, \bibinfo {author} {\bibfnamefont {N.~C.}\ \bibnamefont {Luhmann}},
  \bibinfo {author} {\bibfnamefont {N.}~\bibnamefont {Ito}}, \bibinfo {author}
  {\bibfnamefont {A.}~\bibnamefont {Mase}}, \ and\ \bibinfo {author}
  {\bibfnamefont {S.~G.}\ \bibnamefont {Lee}},\ }\bibfield  {title} {\enquote
  {\bibinfo {title} {{Quasi 3-D ECE Imaging System for Study of MHD
  instabilities in KSTAR}},}\ }\href@noop {} {\bibfield  {journal} {\bibinfo
  {journal} {Review of Scientific Instruments}\ }\textbf {\bibinfo {volume}
  {85}},\ \bibinfo {pages} {11D820} (\bibinfo {year} {2014})}\BibitemShut
  {NoStop}%
\bibitem [{\citenamefont {Oh}\ \emph {et~al.}(2018)\citenamefont {Oh},
  \citenamefont {Yoon}, \citenamefont {Jeon}, \citenamefont {Ko}, \citenamefont
  {Hong}, \citenamefont {Lee}, \citenamefont {Kwon}, \citenamefont {Choi},
  \citenamefont {Park}, \citenamefont {Kwak}, \citenamefont {Kim},
  \citenamefont {Nam}, \citenamefont {Wang}, \citenamefont {Jeong},
  \citenamefont {Park}, \citenamefont {Kim}, \citenamefont {In}, \citenamefont
  {Park}, \citenamefont {Yun}, \citenamefont {Choe}, \citenamefont {Ghim},
  \citenamefont {Na},\ and\ \citenamefont {Hwang}}]{OhJKPS2018}%
  \BibitemOpen
  \bibfield  {author} {\bibinfo {author} {\bibfnamefont {Y.~K.}\ \bibnamefont
  {Oh}}, \bibinfo {author} {\bibfnamefont {S.}~\bibnamefont {Yoon}}, \bibinfo
  {author} {\bibfnamefont {Y.-M.}\ \bibnamefont {Jeon}}, \bibinfo {author}
  {\bibfnamefont {W.-H.}\ \bibnamefont {Ko}}, \bibinfo {author} {\bibfnamefont
  {S.-H.}\ \bibnamefont {Hong}}, \bibinfo {author} {\bibfnamefont {H.-H.}\
  \bibnamefont {Lee}}, \bibinfo {author} {\bibfnamefont {J.-M.}\ \bibnamefont
  {Kwon}}, \bibinfo {author} {\bibfnamefont {M.}~\bibnamefont {Choi}}, \bibinfo
  {author} {\bibfnamefont {B.-H.}\ \bibnamefont {Park}}, \bibinfo {author}
  {\bibfnamefont {J.-G.}\ \bibnamefont {Kwak}}, \bibinfo {author}
  {\bibfnamefont {W.-C.}\ \bibnamefont {Kim}}, \bibinfo {author} {\bibfnamefont
  {Y.-U.}\ \bibnamefont {Nam}}, \bibinfo {author} {\bibfnamefont
  {S.}~\bibnamefont {Wang}}, \bibinfo {author} {\bibfnamefont {J.-H.}\
  \bibnamefont {Jeong}}, \bibinfo {author} {\bibfnamefont {K.-r.}\ \bibnamefont
  {Park}}, \bibinfo {author} {\bibfnamefont {Y.-S.}\ \bibnamefont {Kim}},
  \bibinfo {author} {\bibfnamefont {Y.}~\bibnamefont {In}}, \bibinfo {author}
  {\bibfnamefont {H.~K.}\ \bibnamefont {Park}}, \bibinfo {author}
  {\bibfnamefont {G.}~\bibnamefont {Yun}}, \bibinfo {author} {\bibfnamefont
  {W.}~\bibnamefont {Choe}}, \bibinfo {author} {\bibfnamefont {Y.-C.}\
  \bibnamefont {Ghim}}, \bibinfo {author} {\bibfnamefont {Y.-S.}\ \bibnamefont
  {Na}}, \ and\ \bibinfo {author} {\bibfnamefont {Y.~S.}\ \bibnamefont
  {Hwang}},\ }\bibfield  {title} {\enquote {\bibinfo {title} {{Progress of the
  KSTAR Research Program Exploring the Advanced High Performance and
  Steady-State Plasma Operations}},}\ }\href@noop {} {\bibfield  {journal}
  {\bibinfo  {journal} {Journal of the Korean Physical Society}\ }\textbf
  {\bibinfo {volume} {73}},\ \bibinfo {pages} {712--735} (\bibinfo {year}
  {2018})}\BibitemShut {NoStop}%
\bibitem [{\citenamefont {Bak}, \citenamefont {Lee},\ and\ \citenamefont
  {Son}(2004)}]{BakRSI2004}%
  \BibitemOpen
  \bibfield  {author} {\bibinfo {author} {\bibfnamefont {J.~G.}\ \bibnamefont
  {Bak}}, \bibinfo {author} {\bibfnamefont {S.~G.}\ \bibnamefont {Lee}}, \ and\
  \bibinfo {author} {\bibfnamefont {D.}~\bibnamefont {Son}},\ }\bibfield
  {title} {\enquote {\bibinfo {title} {Performance of the magnetic sensor and
  the integrator for the kstar magnetic diagnostics},}\ }\href {\doibase
  10.1063/1.1789620} {\bibfield  {journal} {\bibinfo  {journal} {Review of
  Scientific Instruments}\ }\textbf {\bibinfo {volume} {75}},\ \bibinfo {pages}
  {4305--4307} (\bibinfo {year} {2004})},\ \Eprint
  {http://arxiv.org/abs/https://doi.org/10.1063/1.1789620}
  {https://doi.org/10.1063/1.1789620} \BibitemShut {NoStop}%
\bibitem [{\citenamefont {Kogi}\ \emph {et~al.}(2010)\citenamefont {Kogi},
  \citenamefont {Jeong}, \citenamefont {Lee}, \citenamefont {Akaki},
  \citenamefont {Mase}, \citenamefont {Kuwahara}, \citenamefont {Yoshinaga},
  \citenamefont {Nagayama}, \citenamefont {Kwon},\ and\ \citenamefont
  {Kawahata}}]{KogiRSI2010}%
  \BibitemOpen
  \bibfield  {author} {\bibinfo {author} {\bibfnamefont {Y.}~\bibnamefont
  {Kogi}}, \bibinfo {author} {\bibfnamefont {S.~H.}\ \bibnamefont {Jeong}},
  \bibinfo {author} {\bibfnamefont {K.~D.}\ \bibnamefont {Lee}}, \bibinfo
  {author} {\bibfnamefont {K.}~\bibnamefont {Akaki}}, \bibinfo {author}
  {\bibfnamefont {A.}~\bibnamefont {Mase}}, \bibinfo {author} {\bibfnamefont
  {D.}~\bibnamefont {Kuwahara}}, \bibinfo {author} {\bibfnamefont
  {T.}~\bibnamefont {Yoshinaga}}, \bibinfo {author} {\bibfnamefont
  {Y.}~\bibnamefont {Nagayama}}, \bibinfo {author} {\bibfnamefont
  {M.}~\bibnamefont {Kwon}}, \ and\ \bibinfo {author} {\bibfnamefont
  {K.}~\bibnamefont {Kawahata}},\ }\bibfield  {title} {\enquote {\bibinfo
  {title} {Calibration of electron cyclotron emission radiometer for kstar},}\
  }\href {\doibase 10.1063/1.3491304} {\bibfield  {journal} {\bibinfo
  {journal} {Review of Scientific Instruments}\ }\textbf {\bibinfo {volume}
  {81}},\ \bibinfo {pages} {10D916} (\bibinfo {year} {2010})},\ \Eprint
  {http://arxiv.org/abs/https://doi.org/10.1063/1.3491304}
  {https://doi.org/10.1063/1.3491304} \BibitemShut {NoStop}%
\bibitem [{\citenamefont {Ko}, \citenamefont {Oh},\ and\ \citenamefont
  {Kwon}(2010)}]{KoIEEE2010}%
  \BibitemOpen
  \bibfield  {author} {\bibinfo {author} {\bibfnamefont {W.~H.}\ \bibnamefont
  {Ko}}, \bibinfo {author} {\bibfnamefont {S.}~\bibnamefont {Oh}}, \ and\
  \bibinfo {author} {\bibfnamefont {M.}~\bibnamefont {Kwon}},\ }\bibfield
  {title} {\enquote {\bibinfo {title} {Kstar charge exchange spectroscopy
  system},}\ }\href {\doibase 10.1109/TPS.2010.2042182} {\bibfield  {journal}
  {\bibinfo  {journal} {IEEE Transactions on Plasma Science}\ }\textbf
  {\bibinfo {volume} {38}},\ \bibinfo {pages} {996--1000} (\bibinfo {year}
  {2010})}\BibitemShut {NoStop}%
\bibitem [{\citenamefont {Sarazin}\ \emph {et~al.}(2011)\citenamefont
  {Sarazin}, \citenamefont {Grandgirard}, \citenamefont {Abiteboul},
  \citenamefont {Allfrey}, \citenamefont {Garbet}, \citenamefont {Ghendrih},
  \citenamefont {Latu}, \citenamefont {Strugarek}, \citenamefont
  {Dif-Pradalier}, \citenamefont {Diamond}, \citenamefont {Ku}, \citenamefont
  {Chang}, \citenamefont {McMillan}, \citenamefont {Tran}, \citenamefont
  {Villard}, \citenamefont {Jolliet}, \citenamefont {Bottino},\ and\
  \citenamefont {Angelino}}]{SarazinNF2011}%
  \BibitemOpen
  \bibfield  {author} {\bibinfo {author} {\bibfnamefont {Y.}~\bibnamefont
  {Sarazin}}, \bibinfo {author} {\bibfnamefont {V.}~\bibnamefont
  {Grandgirard}}, \bibinfo {author} {\bibfnamefont {J.}~\bibnamefont
  {Abiteboul}}, \bibinfo {author} {\bibfnamefont {S.}~\bibnamefont {Allfrey}},
  \bibinfo {author} {\bibfnamefont {X.}~\bibnamefont {Garbet}}, \bibinfo
  {author} {\bibfnamefont {P.}~\bibnamefont {Ghendrih}}, \bibinfo {author}
  {\bibfnamefont {G.}~\bibnamefont {Latu}}, \bibinfo {author} {\bibfnamefont
  {A.}~\bibnamefont {Strugarek}}, \bibinfo {author} {\bibfnamefont
  {G.}~\bibnamefont {Dif-Pradalier}}, \bibinfo {author} {\bibfnamefont {P.~H.}\
  \bibnamefont {Diamond}}, \bibinfo {author} {\bibfnamefont {S.}~\bibnamefont
  {Ku}}, \bibinfo {author} {\bibfnamefont {C.~S.}\ \bibnamefont {Chang}},
  \bibinfo {author} {\bibfnamefont {B.~F.}\ \bibnamefont {McMillan}}, \bibinfo
  {author} {\bibfnamefont {T.~M.}\ \bibnamefont {Tran}}, \bibinfo {author}
  {\bibfnamefont {L.}~\bibnamefont {Villard}}, \bibinfo {author} {\bibfnamefont
  {S.}~\bibnamefont {Jolliet}}, \bibinfo {author} {\bibfnamefont
  {A.}~\bibnamefont {Bottino}}, \ and\ \bibinfo {author} {\bibfnamefont
  {P.}~\bibnamefont {Angelino}},\ }\bibfield  {title} {\enquote {\bibinfo
  {title} {{Predictions on heat transport and plasma rotation from global
  gyrokinetic simulations}},}\ }\href@noop {} {\bibfield  {journal} {\bibinfo
  {journal} {Nuclear Fusion}\ }\textbf {\bibinfo {volume} {51}},\ \bibinfo
  {pages} {103023} (\bibinfo {year} {2011})}\BibitemShut {NoStop}%
\bibitem [{\citenamefont {Hwa}\ and\ \citenamefont
  {Kardar}(1992)}]{HwaPRA1992}%
  \BibitemOpen
  \bibfield  {author} {\bibinfo {author} {\bibfnamefont {T.}~\bibnamefont
  {Hwa}}\ and\ \bibinfo {author} {\bibfnamefont {M.}~\bibnamefont {Kardar}},\
  }\bibfield  {title} {\enquote {\bibinfo {title} {{Avalanches, hydrodynamics,
  and discharge events in models of sandpiles}},}\ }\href@noop {} {\bibfield
  {journal} {\bibinfo  {journal} {Physical Review A}\ }\textbf {\bibinfo
  {volume} {45}},\ \bibinfo {pages} {7002--7023} (\bibinfo {year}
  {1992})}\BibitemShut {NoStop}%
\bibitem [{\citenamefont {Mier}\ \emph {et~al.}(2008)\citenamefont {Mier},
  \citenamefont {Sanchez}, \citenamefont {Garc{\'\i}a}, \citenamefont
  {Newman},\ and\ \citenamefont {Carreras}}]{MierPoP2008}%
  \BibitemOpen
  \bibfield  {author} {\bibinfo {author} {\bibfnamefont {J.~A.}\ \bibnamefont
  {Mier}}, \bibinfo {author} {\bibfnamefont {R.}~\bibnamefont {Sanchez}},
  \bibinfo {author} {\bibfnamefont {L.}~\bibnamefont {Garc{\'\i}a}}, \bibinfo
  {author} {\bibfnamefont {D.~E.}\ \bibnamefont {Newman}}, \ and\ \bibinfo
  {author} {\bibfnamefont {B.~A.}\ \bibnamefont {Carreras}},\ }\bibfield
  {title} {\enquote {\bibinfo {title} {{On the nature of transport in
  near-critical dissipative-trapped-electron-mode turbulence: Effect of a
  subdominant diffusive channel}},}\ }\href@noop {} {\bibfield  {journal}
  {\bibinfo  {journal} {Physics of Plasmas}\ }\textbf {\bibinfo {volume}
  {15}},\ \bibinfo {pages} {112301} (\bibinfo {year} {2008})}\BibitemShut
  {NoStop}%
\bibitem [{\citenamefont {Carreras}\ \emph {et~al.}(1998)\citenamefont
  {Carreras}, \citenamefont {van Milligen}, \citenamefont {Pedrosa},
  \citenamefont {Balb{\'\i}n}, \citenamefont {Hidalgo}, \citenamefont {Newman},
  \citenamefont {S{\'a}nchez}, \citenamefont {Frances}, \citenamefont
  {Garc{\'\i}a-Cort{\'e}s}, \citenamefont {Bleuel}, \citenamefont {Endler},
  \citenamefont {Davies},\ and\ \citenamefont {Matthews}}]{CarrerasPRL1998}%
  \BibitemOpen
  \bibfield  {author} {\bibinfo {author} {\bibfnamefont {B.~A.}\ \bibnamefont
  {Carreras}}, \bibinfo {author} {\bibfnamefont {B.~P.}\ \bibnamefont {van
  Milligen}}, \bibinfo {author} {\bibfnamefont {M.~A.}\ \bibnamefont
  {Pedrosa}}, \bibinfo {author} {\bibfnamefont {R.}~\bibnamefont
  {Balb{\'\i}n}}, \bibinfo {author} {\bibfnamefont {C.}~\bibnamefont
  {Hidalgo}}, \bibinfo {author} {\bibfnamefont {D.~E.}\ \bibnamefont {Newman}},
  \bibinfo {author} {\bibfnamefont {E.}~\bibnamefont {S{\'a}nchez}}, \bibinfo
  {author} {\bibfnamefont {M.}~\bibnamefont {Frances}}, \bibinfo {author}
  {\bibfnamefont {I.}~\bibnamefont {Garc{\'\i}a-Cort{\'e}s}}, \bibinfo {author}
  {\bibfnamefont {J.}~\bibnamefont {Bleuel}}, \bibinfo {author} {\bibfnamefont
  {M.}~\bibnamefont {Endler}}, \bibinfo {author} {\bibfnamefont
  {S.}~\bibnamefont {Davies}}, \ and\ \bibinfo {author} {\bibfnamefont {G.~F.}\
  \bibnamefont {Matthews}},\ }\bibfield  {title} {\enquote {\bibinfo {title}
  {{Long-Range Time Correlations in Plasma Edge Turbulence}},}\ }\href@noop {}
  {\bibfield  {journal} {\bibinfo  {journal} {Physical Review Letters}\
  }\textbf {\bibinfo {volume} {80}},\ \bibinfo {pages} {4438--4441} (\bibinfo
  {year} {1998})}\BibitemShut {NoStop}%
\bibitem [{\citenamefont {Politzer}\ \emph {et~al.}(2002)\citenamefont
  {Politzer}, \citenamefont {Austin}, \citenamefont {Gilmore}, \citenamefont
  {McKee}, \citenamefont {Rhodes}, \citenamefont {Yu}, \citenamefont {Doyle},
  \citenamefont {Evans},\ and\ \citenamefont {Moyere}}]{PolitzerPoP2002}%
  \BibitemOpen
  \bibfield  {author} {\bibinfo {author} {\bibfnamefont {P.~A.}\ \bibnamefont
  {Politzer}}, \bibinfo {author} {\bibfnamefont {M.~E.}\ \bibnamefont
  {Austin}}, \bibinfo {author} {\bibfnamefont {M.}~\bibnamefont {Gilmore}},
  \bibinfo {author} {\bibfnamefont {G.~R.}\ \bibnamefont {McKee}}, \bibinfo
  {author} {\bibfnamefont {T.~L.}\ \bibnamefont {Rhodes}}, \bibinfo {author}
  {\bibfnamefont {C.~X.}\ \bibnamefont {Yu}}, \bibinfo {author} {\bibfnamefont
  {E.~J.}\ \bibnamefont {Doyle}}, \bibinfo {author} {\bibfnamefont {T.~E.}\
  \bibnamefont {Evans}}, \ and\ \bibinfo {author} {\bibfnamefont {R.~A.}\
  \bibnamefont {Moyere}},\ }\bibfield  {title} {\enquote {\bibinfo {title}
  {{Characterization of avalanche-like events in a confined plasma}},}\
  }\href@noop {} {\bibfield  {journal} {\bibinfo  {journal} {Physics of
  Plasmas}\ }\textbf {\bibinfo {volume} {9}},\ \bibinfo {pages} {1962--1969}
  (\bibinfo {year} {2002})}\BibitemShut {NoStop}%
\bibitem [{\citenamefont {Chung}\ \emph {et~al.}(2017)\citenamefont {Chung},
  \citenamefont {Kim}, \citenamefont {Jeon}, \citenamefont {Kim}, \citenamefont
  {Choi}, \citenamefont {Ko}, \citenamefont {Lee}, \citenamefont {Lee},
  \citenamefont {Yi}, \citenamefont {Kwon}, \citenamefont {Hahn}, \citenamefont
  {Ko}, \citenamefont {Lee},\ and\ \citenamefont {Yoon}}]{ChungNF2017}%
  \BibitemOpen
  \bibfield  {author} {\bibinfo {author} {\bibfnamefont {J.}~\bibnamefont
  {Chung}}, \bibinfo {author} {\bibfnamefont {H.~S.}\ \bibnamefont {Kim}},
  \bibinfo {author} {\bibfnamefont {Y.~M.}\ \bibnamefont {Jeon}}, \bibinfo
  {author} {\bibfnamefont {J.}~\bibnamefont {Kim}}, \bibinfo {author}
  {\bibfnamefont {M.~J.}\ \bibnamefont {Choi}}, \bibinfo {author}
  {\bibfnamefont {J.}~\bibnamefont {Ko}}, \bibinfo {author} {\bibfnamefont
  {K.~D.}\ \bibnamefont {Lee}}, \bibinfo {author} {\bibfnamefont {H.~H.}\
  \bibnamefont {Lee}}, \bibinfo {author} {\bibfnamefont {S.}~\bibnamefont
  {Yi}}, \bibinfo {author} {\bibfnamefont {J.~M.}\ \bibnamefont {Kwon}},
  \bibinfo {author} {\bibfnamefont {S.~H.}\ \bibnamefont {Hahn}}, \bibinfo
  {author} {\bibfnamefont {W.~H.}\ \bibnamefont {Ko}}, \bibinfo {author}
  {\bibfnamefont {J.~H.}\ \bibnamefont {Lee}}, \ and\ \bibinfo {author}
  {\bibfnamefont {S.~W.}\ \bibnamefont {Yoon}},\ }\bibfield  {title} {\enquote
  {\bibinfo {title} {{Formation of the internal transport barrier in KSTAR}},}\
  }\href@noop {} {\bibfield  {journal} {\bibinfo  {journal} {Nuclear Fusion}\
  }\textbf {\bibinfo {volume} {58}},\ \bibinfo {pages} {016019} (\bibinfo
  {year} {2017})}\BibitemShut {NoStop}%
\bibitem [{\citenamefont {Takeji}\ \emph {et~al.}(1998)\citenamefont {Takeji},
  \citenamefont {Kamada}, \citenamefont {Ozeki}, \citenamefont {Ishida},
  \citenamefont {Takizuka}, \citenamefont {Neyatani},\ and\ \citenamefont
  {Tokuda}}]{TakejiPoP1998}%
  \BibitemOpen
  \bibfield  {author} {\bibinfo {author} {\bibfnamefont {S.}~\bibnamefont
  {Takeji}}, \bibinfo {author} {\bibfnamefont {Y.}~\bibnamefont {Kamada}},
  \bibinfo {author} {\bibfnamefont {T.}~\bibnamefont {Ozeki}}, \bibinfo
  {author} {\bibfnamefont {S.}~\bibnamefont {Ishida}}, \bibinfo {author}
  {\bibfnamefont {T.}~\bibnamefont {Takizuka}}, \bibinfo {author}
  {\bibfnamefont {Y.}~\bibnamefont {Neyatani}}, \ and\ \bibinfo {author}
  {\bibfnamefont {S.}~\bibnamefont {Tokuda}},\ }\bibfield  {title} {\enquote
  {\bibinfo {title} {{Ideal magnetohydrodynamic instabilities with low toroidal
  mode numbers localized near an internal transport barrier in high-$\beta$p
  mode plasmas in the Japan Atomic Energy Research Institute Tokamak-60
  Upgrade}},}\ }\href@noop {} {\bibfield  {journal} {\bibinfo  {journal}
  {Physics of Plasmas}\ }\textbf {\bibinfo {volume} {4}},\ \bibinfo {pages}
  {4283--4291} (\bibinfo {year} {1998})}\BibitemShut {NoStop}%
\bibitem [{\citenamefont {Manickam}\ \emph {et~al.}(1999)\citenamefont
  {Manickam}, \citenamefont {Fujita}, \citenamefont {Gorelenkov}, \citenamefont
  {Isayama}, \citenamefont {Kamada}, \citenamefont {Okabayashi}, \citenamefont
  {Bell}, \citenamefont {Bell}, \citenamefont {Budny}, \citenamefont
  {Fredrickson}, \citenamefont {Ishida}, \citenamefont {Ishii}, \citenamefont
  {Levinton}, \citenamefont {Ozeki}, \citenamefont {Shirai}, \citenamefont
  {Takeji},\ and\ \citenamefont {Zarnstorff}}]{ManickamNF1999}%
  \BibitemOpen
  \bibfield  {author} {\bibinfo {author} {\bibfnamefont {J.}~\bibnamefont
  {Manickam}}, \bibinfo {author} {\bibfnamefont {T.}~\bibnamefont {Fujita}},
  \bibinfo {author} {\bibfnamefont {N.}~\bibnamefont {Gorelenkov}}, \bibinfo
  {author} {\bibfnamefont {A.}~\bibnamefont {Isayama}}, \bibinfo {author}
  {\bibfnamefont {Y.}~\bibnamefont {Kamada}}, \bibinfo {author} {\bibfnamefont
  {M.}~\bibnamefont {Okabayashi}}, \bibinfo {author} {\bibfnamefont
  {M.}~\bibnamefont {Bell}}, \bibinfo {author} {\bibfnamefont {R.}~\bibnamefont
  {Bell}}, \bibinfo {author} {\bibfnamefont {R.}~\bibnamefont {Budny}},
  \bibinfo {author} {\bibfnamefont {E.}~\bibnamefont {Fredrickson}}, \bibinfo
  {author} {\bibfnamefont {S.}~\bibnamefont {Ishida}}, \bibinfo {author}
  {\bibfnamefont {Y.}~\bibnamefont {Ishii}}, \bibinfo {author} {\bibfnamefont
  {F.}~\bibnamefont {Levinton}}, \bibinfo {author} {\bibfnamefont
  {T.}~\bibnamefont {Ozeki}}, \bibinfo {author} {\bibfnamefont
  {H.}~\bibnamefont {Shirai}}, \bibinfo {author} {\bibfnamefont
  {S.}~\bibnamefont {Takeji}}, \ and\ \bibinfo {author} {\bibfnamefont
  {M.}~\bibnamefont {Zarnstorff}},\ }\bibfield  {title} {\enquote {\bibinfo
  {title} {{Localized MHD activity near internal transport barriers in JT-60U
  and TFTR}},}\ }\href@noop {} {\bibfield  {journal} {\bibinfo  {journal}
  {Nuclear Fusion}\ }\textbf {\bibinfo {volume} {39}},\ \bibinfo {pages}
  {1819--1826} (\bibinfo {year} {1999})}\BibitemShut {NoStop}%
\bibitem [{\citenamefont {Candy}\ and\ \citenamefont
  {Waltz}(2003)}]{CandyPRL2003}%
  \BibitemOpen
  \bibfield  {author} {\bibinfo {author} {\bibfnamefont {J.}~\bibnamefont
  {Candy}}\ and\ \bibinfo {author} {\bibfnamefont {R.~E.}\ \bibnamefont
  {Waltz}},\ }\bibfield  {title} {\enquote {\bibinfo {title} {{Anomalous
  Transport Scaling in the DIII-D Tokamak Matched by Supercomputer
  Simulation}},}\ }\href@noop {} {\bibfield  {journal} {\bibinfo  {journal}
  {Physical Review Letters}\ }\textbf {\bibinfo {volume} {91}},\ \bibinfo
  {pages} {A281} (\bibinfo {year} {2003})}\BibitemShut {NoStop}%
\bibitem [{\citenamefont {Dif-Pradalier}\ \emph {et~al.}(2017)\citenamefont
  {Dif-Pradalier}, \citenamefont {Hornung}, \citenamefont {Garbet},
  \citenamefont {Ghendrih}, \citenamefont {Grandgirard}, \citenamefont {Latu},\
  and\ \citenamefont {Sarazin}}]{DifNF2017}%
  \BibitemOpen
  \bibfield  {author} {\bibinfo {author} {\bibfnamefont {G.}~\bibnamefont
  {Dif-Pradalier}}, \bibinfo {author} {\bibfnamefont {G.}~\bibnamefont
  {Hornung}}, \bibinfo {author} {\bibfnamefont {X.}~\bibnamefont {Garbet}},
  \bibinfo {author} {\bibfnamefont {P.}~\bibnamefont {Ghendrih}}, \bibinfo
  {author} {\bibfnamefont {V.}~\bibnamefont {Grandgirard}}, \bibinfo {author}
  {\bibfnamefont {G.}~\bibnamefont {Latu}}, \ and\ \bibinfo {author}
  {\bibfnamefont {Y.}~\bibnamefont {Sarazin}},\ }\bibfield  {title} {\enquote
  {\bibinfo {title} {{The E $\times$ B staircase of magnetised plasmas}},}\
  }\href@noop {} {\bibfield  {journal} {\bibinfo  {journal} {Nuclear Fusion}\
  }\textbf {\bibinfo {volume} {57}},\ \bibinfo {pages} {066026} (\bibinfo
  {year} {2017})}\BibitemShut {NoStop}%
\bibitem [{\citenamefont {Idomura}(2016)}]{IdomuraJCP2016}%
  \BibitemOpen
  \bibfield  {author} {\bibinfo {author} {\bibfnamefont {Y.}~\bibnamefont
  {Idomura}},\ }\bibfield  {title} {\enquote {\bibinfo {title} {{A new hybrid
  kinetic electron model for full-f gyrokinetic simulations}},}\ }\href@noop {}
  {\bibfield  {journal} {\bibinfo  {journal} {Journal of Computational
  Physics}\ }\textbf {\bibinfo {volume} {313}},\ \bibinfo {pages} {511--531}
  (\bibinfo {year} {2016})}\BibitemShut {NoStop}%
\bibitem [{\citenamefont {Lee}\ \emph {et~al.}(2018)\citenamefont {Lee},
  \citenamefont {Leem}, \citenamefont {Lee}, \citenamefont {Choi},
  \citenamefont {Park}, \citenamefont {Lee}, \citenamefont {Yun}, \citenamefont
  {Kim}, \citenamefont {Park}, \citenamefont {Kim},\ and\ \citenamefont {{the
  KSTAR team}}}]{LeePPCF2018}%
  \BibitemOpen
  \bibfield  {author} {\bibinfo {author} {\bibfnamefont {W.}~\bibnamefont
  {Lee}}, \bibinfo {author} {\bibfnamefont {J.}~\bibnamefont {Leem}}, \bibinfo
  {author} {\bibfnamefont {D.~J.}\ \bibnamefont {Lee}}, \bibinfo {author}
  {\bibfnamefont {M.~J.}\ \bibnamefont {Choi}}, \bibinfo {author}
  {\bibfnamefont {H.~K.}\ \bibnamefont {Park}}, \bibinfo {author}
  {\bibfnamefont {J.~a.}\ \bibnamefont {Lee}}, \bibinfo {author} {\bibfnamefont
  {G.~S.}\ \bibnamefont {Yun}}, \bibinfo {author} {\bibfnamefont {T.~G.}\
  \bibnamefont {Kim}}, \bibinfo {author} {\bibfnamefont {H.}~\bibnamefont
  {Park}}, \bibinfo {author} {\bibfnamefont {K.~W.}\ \bibnamefont {Kim}}, \
  and\ \bibinfo {author} {\bibnamefont {{the KSTAR team}}},\ }\bibfield
  {title} {\enquote {\bibinfo {title} {{Quasi-coherent fluctuation measurement
  with the upgraded microwave imaging reflectometer in KSTAR}},}\ }\href@noop
  {} {\bibfield  {journal} {\bibinfo  {journal} {Plasma Physics and Controlled
  Fusion}\ }\textbf {\bibinfo {volume} {60}},\ \bibinfo {pages} {115009}
  (\bibinfo {year} {2018})}\BibitemShut {NoStop}%
\end{thebibliography}

%

\end{document}